\documentclass{imsart}
\usepackage{amsthm}
\usepackage{bigints}
\usepackage{amsmath}
\usepackage{natbib}
\usepackage[colorlinks,citecolor=blue,urlcolor=blue,filecolor=blue,backref=page]{hyperref}
\usepackage{graphicx}
\usepackage{amsfonts,amssymb,amsmath,amscd,amsthm,latexsym}
\usepackage{natbib}
\usepackage{graphicx}
\usepackage[]{units}

\usepackage[english]{babel}
\usepackage[utf8]{inputenc}
\usepackage{amsmath}
\usepackage{amsfonts}
\usepackage{graphicx}
\usepackage{algorithm}
\usepackage{algpseudocode}

\usepackage[usenames,dvipsnames,svgnames,table]{xcolor}

\usepackage{algorithm}
\usepackage{algpseudocode}
\usepackage{qtree}
\usepackage{xcolor}
\usepackage{comment}

\usepackage{times}
\usepackage{bm}
\usepackage{natbib}
\usepackage{amssymb}
\usepackage{graphicx, enumerate}

\newcommand{\erre}{{\ttfamily{{\bf R}\,}}}

\newtheorem{exemp}{Example}[section]

\startlocaldefs
\endlocaldefs

\begin{document}


\begin{frontmatter}
\title{Approximate Bayesian inference in semiparametric copula models\protect\thanksref{T1}}

\runtitle{Approximate Bayesian inference in semiparametric copula models}
\thankstext{T1}{Clara Grazian, Nuffield Department of Medicine, University of Oxford, Old Road Campus, Roosevelt Dr, Oxford OX3 7FZ, United Kingdom,
{\sf clara.grazian@ndm.ox.ac.uk}, and Brunero Liseo, MEMOTEF, Sapienza  Universit{\` a} di Roma, 
via del Castro Laurenziano 9, 00161, Italy {\sf brunero.liseo@uniroma1.it}. }

\begin{aug}
\author{\fnms{Clara} \snm{Grazian}\thanksref{addr1}}
\and
\author{\fnms{Brunero} \snm{Liseo}\thanksref{addr2}}

\runauthor{}

\address[addr1]{Nuffield Department of Medicine, University of Oxford}
\address[addr2]{MEMOTEF, Sapienza Universit\`a di Roma}


\end{aug}

\begin{abstract}
We describe a simple method for making inference on a functional of a multivariate distribution, based on its copula representation. We make use of an approximate Bayesian Monte\,Carlo algorithm, where the proposed values of the functional of interest 
are weighted in terms of their Bayesian exponentially tilted empirical likelihood.
This method is particularly useful when the ``true'' likelihood function associated with 
the working model is too costly to evaluate or when the working model is only partially specified.
\end{abstract}


\end{frontmatter}






\section{Introduction}
\label{sec:1}
Copula models are widely used in multivariate data analysis. 
Major areas of application include econometrics \citep{spri}, geophysics \citep{geop}, 
climate prediction \citep{clima}, actuarial science and finance \citep{cheru}, among the others.
A copula allows a useful representation of the joint distribution of a random vector in two steps: 
the marginal distributions and a distribution function which captures the dependence among the vector components.  

From a statistical perspective, whereas it is generally simple to produce reliable 
estimates of the parameters of the marginal 
distributions of the data, the problem of estimating the dependence structure, 
however it is modelled, is crucial and complex, especially in high dimensional situations. 
A list of important applications can be found in the recent 
monograph by \cite{joe}.

In a frequentist approach to copula models, there are no broadly satisfactory methods for the 
joint estimation of marginal and copula parameters. The most popular method is the so called 
\textit{Inference Functions for Margins} method, where the parameters of the marginal distributions 
are estimated first, and then pseudo-data are obtained by plugging-in the estimates 
of the marginal parameters. Then inference on the copula parameters is performed using the pseudo-data: this approach
does not account for the uncertainty on the estimation of the marginal parameters. A nonparametric alternative may be found in \cite{kauermann:schellhase:ruppert:13}, where a penalized hierarchical B-splines approach is proposed.

The literature on Bayesian alternatives is still limited, although they show a great potential for inference in a number of cases, for example in the modelling of multivariate discrete data \citep{smith:khaled:12} and of conditional copulae \citep{craiu:12}; see \cite{smith2} for a review on parametric methods and \cite{wu:14} for a nonparametric approach. An instrumental use of copulas in Bayesian mixture estimation may be found in \cite{burda:prokhorov:14}. An alternative and more flexible copula construction is based on the so-called vines, where the joint dependence structure of a multivariate random vector is decomposed into several marginal and conditional bivariate copulas. A Bayesian use of this approach can be found in \cite{min:10} and \cite{gruber:czado:15}.

In this work we consider the general problem of estimating a functional of interest of a generic copula: practical illustrations will include the Spearman's $\rho$ and tail dependence indices. 
Our method is based on the simulation of a posterior sample weighted in terms of the Bayesian exponentially tilted empirical likelihood \citep{schen:05}. 
A similar approach, in a frequentist fashion, has been proposed in \cite{oh:13}, where a simulated method of moments is discussed for copula estimation. The main difference between the two approaches is that in our method the functional represents the actual quantity of interest and no assumption is made on the copula structure, while in \cite{oh:13} the functional is only instrumental, as a moment condition, to estimate the parameter of a given parametric copula. 

As already stated, the central tool in our approach is the empirical likelihood  \citep{owen}; we adopt an approximate Bayesian approach based on the use of a pseudo-likelihood, along the lines of \cite{abcel}. 
We use a partially specified model where the prior distribution is explicitly elicited only on the quantity of interest. Its approximate posterior distribution is obtained via the use of the Bayesian exponentially tilted empirical likelihood approximation of the marginal likelihood of 
the quantity of interest, illustrated in \cite{schen:05}.
This approximation of the true ``unknown'' likelihood function hopefully reduces the potential bias for incorrect distributional assumptions, very hard to check in complex dependence modeling.
Our approach can be adapted both to parametric and nonparametric modeling of the marginal distributions.

A brief review on copula models and empirical likelihood methods will be given in Section \ref{sec:preliminaries}. The method used to approximate the posterior distribution for a functional of the copula will be presented in Section \ref{sec:proposal} and its asymptotical justification will be studied in Section \ref{sec:asymptotics}. The rest of the paper is devoted to the illustration of the empirical behaviour of our proposal on simulated and real data sets; in particular, Section \ref{sec:srho} will apply the method for the Bayesian estimation of the Spearman's $\rho$ in a bivariate setting, while Section \ref{sec:TD} will be focused on tail depedence coefficients; multivariate extensions will be available in Section \ref{sec:multianalysis}. Finally, Section \ref{sec:logreturns} contains a real financial application. A discussion concludes the work. 

\section{Preliminaries: Copulae and Empirical Likelihood}
\label{sec:preliminaries}

A copula represents an alternative way of writing the joint distribution of a random vector
$X= (X_1, \dots, X_d)$. Given a $d$-variate cumulative distribution function $F$ which depends on some parameter $\theta$, 
it is possible to show \citep{sklar} that there always exists a $d$-variate function $C_{\theta} : [0, 1]^d \to [0, 1]$, such that

$$F(x_1,\dots , x_d; \lambda ,\theta) = C_{\theta}(F_1 (x_1;\lambda_1),\dots,  F_d (x_d;\lambda_d)),$$ 

\noindent where $\lambda = (\lambda_1, \dots, \lambda_d)$ is a vector of parameters, $F_j$ is the marginal distribution of $X_j$ depending on a parameter vector $\lambda_j$.
In other terms, the copula $C$ is a distribution function with uniform margins on $[0, 1]$: it binds together the univariate $F_1, F_2, \dots , F_d$ in order to produce the $d$-variate distribution $F$.
The copula $C$ does not depend on the marginal distributions, and it 
accounts for potential dependence among the components of the random vector $X$.

In the continuous case, the density of a random vector $(X_1, \dots X_d)$ has a unique copula representation given by

\begin{equation}
\label{eq:copdens}
f(x|\lambda,\theta) = c(u;\theta) \prod_{j=1}^d f_j(x_j|\lambda_j)
\end{equation}

\noindent where $u=(u_{1},\cdots,u_{d})=(F_1(x_{1};\lambda_1),\cdots,F_d(x_{d};\lambda_d))$, $c(u;\theta)$ is the derivative of $C_\theta$ and $\theta$ and $(\lambda_1,\cdots,\lambda_d)$ are the parameter of the copula and of the marginal density functions respectively. 

Given a prior $\pi(\theta,\lambda_1,\cdots,\lambda_d)$ and a sample of size $n$ of independent multivariate observations $(x_{i1},\cdots,x_{id})$ for $i=1,\cdots,n$, the resulting posterior distribution for the parameter vector is 

$$
\pi(\theta, \lambda | x) \propto \pi(\theta, \lambda) \prod_{i=1}^n\left[  c(u_i;\theta) \prod_{j=1}^d f(x_{ij};\lambda_j) \right].
$$ 

\noindent
Notice that the likelihood function is not separable in $\lambda_1,\cdots,\lambda_d$ and $\theta$ because the $u_{i}$'s depend on the marginal parameter $\lambda$.

In the parametric case, frequentist methods of estimation are generally based either on the simultaneous maximization of the likelihood function in $\theta$ and $\lambda$ or on the so-called method of inference functions for margins (IFM) \citep{joe}: here a maximum likelihood estimate of $\lambda$ is obtained using only the second factor of \eqref{eq:copdens}; then the estimate is plugged into the first factor and an estimate of $\theta$ is based on pseudo-data $\hat{u}_{ij}=F(x_{ij}; \hat{\lambda}_j)$. 
The two methods are not equivalent in general \citep{choros:ibragimov:permiakova:10}. 

The first-step estimation may be performed both parametrically and nonparametrically: \cite{genest95} propose a semiparametric approach where nonparametric estimates are contemplated for the marginals and a specific copula function is used. The proposed estimator is shown to be consistent and asymptotically normal. 

It is possible to modify the two-step procedure of \cite{joe:05} within a Bayesian framework, where the joint posterior distribution $\pi(\theta, \lambda_1,\cdots,\lambda_d | x)$ is evaluated through a Monte Carlo algorithm, with $\theta$ and $(\lambda_1,\cdots,\lambda_d)$ generated separately in a Gibbs sampling scheme; see \cite{pitt:chan:kohn:06} for a discussion. \cite{smith2} provides a review on sampling schemes and possible prior distributions, in particular in the case of Gaussian copula models, both in a continuous and in a discrete setting. 

A Bayesian nonparametric approach is followed by \cite{wu:14}, who model and estimate only the copula density function by using infinite mixture models and treat the marginals as given. In particular, they focus on a mixture of multivariate skew-normal copulas, in order to circumvent the symmetry limitation of the Gaussian copula and to preserve the simplicity of the estimation, which is not the case with the skewed \textit{t} copula. The MCMC implementation follows \cite{kalli:griffin:walker:11}.

In all these cases, the goal of the analysis is about the complete dependence structure. 
The aim of this work is different. First, we allow the marginal distributions $F_j$'s to follow either a parametric or a non parametric model. Secondly, we do not make any parametric assumption for the copula function $C$. Rather, we limit our goal to the estimation of a given functional of interest of $C$, say $\phi(C)$. 
In this respect, we adopt a semiparametric Bayesian strategy for estimating  $\phi(C)$ where the parameter of interest is the particular functional $\phi$ for which we derive an approximated posterior distribution 
$$
\pi(\phi|x) \propto \pi(\phi) \hat{L} (\phi; x),
$$

\noindent where $\hat{L}(\phi; x)$ is a nonparametric approximation of the likelihood function for $\phi$. In particular here we use the Bayesian exponentially tilted empirical likelihood of \cite{schen:05}.

We also propose a modified version of the algorithm of \cite{abcel} in a situation 
where the statistical model is only partially specified and the main goal is the 
estimation of a finite dimensional quantity of interest.
In practice this represents the prototypical semiparametric set-up, 
where one is mainly interested in some
characteristics of the population, although the statistical model may contain nuisance parameters 
which are introduced in order to produce more flexible models that might better fit the data at hand. 
In order to make robust inference on the quantity of interest, a reasonable model should account for the uncertainty on the nuisance parameters, in some way.
Even if some of these additional parameters are not particularly important in terms of estimation - they often lack of a precise physical meaning - 
their estimates can dramatically affect inference on the parameter of interest.
In these circumstances it might be more reasonable and robust to partially specify the model. 

Empirical likelihood has been introduced by \cite{owen}; 
it is a way of producing a nonparametric likelihood for a quantity
of interest in an otherwise unspecified statistical model. 
\cite{schen:05} proposed an exponentially tilted empirical likelihood which can also be interpreted as a semiparametric Bayesian procedure. 
Assume that our dataset is composed of $n$ independent
replicates $(x_1,\dots , x_d)$ of some random vector $X$ with
distribution $F$ and corresponding density $f$. 
Rather than defining the usual likelihood function in terms of $f$,
the Bayesian exponentially tilted empirical likelihood is constructed with respect to a given quantity 
of interest, say $\phi$, expressed as a functional of $F$, i.e. $\phi(F)$, 
and then a sort of profile likelihood of $\phi$ is computed
in a nonparametric way. More precisely,
consider a given set of generalized moment conditions of the form
\begin{equation}
E_{F} \left (h(X, \phi) \right )  = 0,
\label{mo-cond}
\end{equation}
where $h(\cdot)$ is a known function, and $\phi$ is the quantity of interest. The resulting Bayesian exponentially tilted empirical likelihood $L_{BEL} (\phi ; x)$ is defined as 

\begin{equation}
\label{eq:lbel}
L_{BEL}(\phi, x) = \prod_{i=1}^n p_i^*(\phi),
\end{equation}

\noindent where $(p_1^*(\phi),\cdots,p_n^*(\phi))$ is the solution of 

\begin{equation*}
\max_{(p_1, \dots, p_n)} \sum_{i=1}^n \left ( -  p_i \log p_i\right ),
\end{equation*}

\noindent under the constraints $0 \leq p_i \leq 1,$  $\sum_{i=1}^n p_i=1$, and 
$\sum_{i=1}^n h(x_i, \phi) p_i = 0.$
The third condition induces
a profiling of the information towards the quantity of interest, through an unbiasedness
condition. 

 \cite{abcel}  proposed a sort of ``sampling importance re-sampling'' \citep{rubin:88} method for dealing with situations where the ``true likelihood'' evaluation is out of reach and parameter values are ``weighted''  by the empirical likelihood proposed by \cite{owen}. Here we replace the empirical likelihood with the exponentially tilted empirical likelihood proposed by \cite{schen:05}, in order to guarantee a solid Bayesian justification of the procedure.
 
\section{The Bayesian Use of Exponentially Tilted Empirical Likelihood }
\label{sec:proposal}

We now describe in detail our method. 
First we illustrate the role played by the two steps, then we present the general algorithm written in a pseudo-code style and we finally comment on the main issues.

\subsection{Step 1: Marginal estimation}
\label{3.1}
We assume that a data set is available in the form of a 
$n\times d$ matrix $x=\left (x_1, x_2, \dots, x_d\right )$,
where $n$ is the sample size and $d$ is the number of variables. Given that the object of interest is a functional of the copula structure, inference for the marginals is not central in this description: one can either use a parametric or a nonparametric model for the marginals.  

In the first case, for each $j=1, \dots, d$, we use data available for $X_{j}$ to derive an estimate of $\pi(\lambda_j|x_j)$; for example, we can generate a sample $\lambda_j=(\lambda_j^{(1)}, \lambda_j^{(2)}, \dots \lambda_j^{(S_j)})$ which is an approximation of the posterior distribution for $\lambda_j$. We allow $S_j$ to vary for $j=1,\cdots,d$ in order to take into account particular features of the marginal models or the information available for each variable $X_j$. 

Alternatively, Bayesian nonparametric estimates of the marginal distributions may be obtained, see \cite{hjort2010} for a general review. In Section \ref{sec:asymptotics} we will argue that the nonparametric choice may lead to better convergence results of the estimation procedure for the functional $\phi$.

\subsection{Step 2: Joint estimation}
\label{3.2}
We also assume that the main focus of the analysis is the estimation of a specific function $\phi$ of $C$; because of this, we avoid to choose the complete copula structure, in order to prevent estimation biases due to model miss-specification, as we will see in Section \ref{sec:srho}.
This is particularly useful and meaningful in those situations
where there is no theoretical or empirical evidence that a given parametric copula should be preferred and we are mainly interested in a synthetic measure of dependence, like for example, the upper tail dependence index between 
two components of $X$, discussed in Section \ref{sec:TD}.
Another popular quantity, which we will consider in Section \ref{sec:srho}, is the Spearman's $\rho$ between two components of $X$, and its extensions to the multivariate case (Section \ref{sec:multianalysis}).
From this perspective, the problems analysed in this paper belong to the class of Bayesian semiparametric problems, where the posterior distribution of a quantity of interest, $\phi$, is investigated, whereas the complete form of the model is considered a nuisance parameter, for which a nonparametric approach seems more cautious and reasonable.

Then, after the estimation of the marginal distributions, performed in Step 1, we now use a copula representation of the multivariate distribution in order to approximate the posterior distribution $\pi(\phi|x)$ of $\phi$, a single measure of the multivariate dependence structure of $X$. 
The posterior distribution of $\phi$ is approximated by combining its prior distribution with the Bayesian Exponentially tilted Empirical Likelihood \eqref{eq:lbel},
$$
\pi(\phi|x) \propto \pi(\phi) L_{BEL} (\phi; x).
$$
As already stated, the Bayesian exponentially tilted empirically likelihood has been introduced by \cite{schen:05}, however its use has been limited so far, with the remarkable exceptions of \cite{lancaster:jae:10} and \cite{yang:he:12} for quantile regression. 

This approach also implies the introduction of a nonparametric prior distribution on the nuisance aspects of the model, as detailed in  \cite{schen:05} and discussed in Section \ref{sec:asymptotics}.  $L_{BEL}(\cdot)$ is computed for a given choice of moment conditions of the form \eqref{mo-cond}, based on a nonparametric estimator of the functional of interest, for which there must exist an (at least, asymptotically) unbiased estimator $\hat{\phi}_n$,  i.e. such that 
\begin{equation*}
E_{F} \left ( \hat{\phi}_n  - \phi \right)  = 0.
\end{equation*}
The existence of an unbiased estimator is a caveat on the use of empirical likelihoods: the problem might be partially circumvented through the use of a bootstrap likelihood as in \cite{bl-leisen}.

\subsection{The algorithm}
\label{3.3}

Here we present the pseudo-code of the proposed algorithm, which we briefly indicate as ABSCop (Approximate Bayesian semiparametric copula). 
In Algorithm \ref{ABCOP} the method is described in the case when parametric models are assumed for the marginals. Nevertheless it could be easily modified to manage the case of nonparametric estimation of the marginal densities, or to a mix of the two cases. 

The final output is a posterior sample drawn from an approximation of the posterior distribution of the quantity of interest $\phi$.

\begin{algorithm}
\caption{ABSCop algorithm}
\label{ABCOP}

	\begin{enumerate}
		\item STEP 1: Marginal estimation \\
		Given a sample $X=(X_1,X_2,\cdots,X_d)$
		 				where $X_j=(X_{1j},\cdots,X_{nj})$ for 
		 				$j=(1,\cdots,d)$ with joint cdf $F_X(x)$ 
		 				and marginal cdf's 
		 				$F_1(x_1;\lambda_1),\cdots,F_d(x_d;\lambda_d)$
		 				
		\begin{algorithmic}[1]
			\For {$j=1,\cdots,d$}
				\State Derive a posterior sample for $\lambda_j$: $(\lambda_j^{1},\cdots,\lambda_j^{S_j})$ approximating the marginal posterior $\pi(\lambda_j|x_j)$
			\EndFor
		\end{algorithmic}
		
		\item STEP 2: Joint estimation

		\begin{algorithmic}[1]
			\For {$b=1,\cdots,B$}
				\State Draw $\phi^{(b)} \sim \pi(\phi)$
				\State Sample one value $\lambda^{s_j}$ from each
							marginal posterior sample: 
							$\lambda'=(\lambda_1^{(s_1)},\cdots,
							\lambda_d^{(s_d)})$
				\State Derive a matrix of uniformly distributed pseudo-data $u_{ij}=F_j(x_{ij};\lambda_j^{(s_j)})$
				$$
u'= \begin{pmatrix}
u_{11}^{(s_1)} & u_{12}^{(s_2)} & \dots & u_{1d}^{(s_d)} \\
u_{21}^{(s_1)} & u_{22}^{(s_2)} & \dots & u_{2d}^{(s_d)} \\
\dots  & \dots  & u_{ij}^{(s_j)} & \dots  \\
u_{n1}^{(s_1)} & u_{n2}^{(s_2)} & \dots & u_{nd}^{(s_d)} \\
\end{pmatrix} .
$$
				\State Compute $L_{BEL}
						(\phi^{(b)};u')=\omega_b$
			\EndFor
			\State \Return A weighted sample of size $B$ of values for $\phi$, where the weights are defined as the $L_{BEL}$, given the nonparametric estimate $\hat{\phi}_n$. 
			\State Sample with replacement $(\phi^{(b)}\omega_b)$, $b=1,\cdots,B$.
		\end{algorithmic}

	\end{enumerate}

\textbf{Output:} a sample of size $B$ of values approximately from the posterior distribution of $\phi$.
\end{algorithm}

There are several critical issues both in the practical implementation of the method and in its theoretical properties, which we will now discuss.

\textbf{Marginal estimation.}\\
As we will see in the real data application in Section \ref{sec:logreturns}, as long as one uses a reasonable parametric model for the marginals, the posterior distribution of $\phi$ will not be seriously affected by this part of model choice. 

\textbf{Weighting the prior sample.}
An interesting point to discuss is that the posterior sample sizes used to approximate the marginal parameter posterior distribution may be different, maybe because estimation of some of them need to be more accurate. Therefore values $S_j$'s are allowed to change across $j=1,\cdots,d$. 

If, instead, all the $S_j = S$ for $j=1,\cdots,d$, it would be ideally possible to run the second step (for each $b=1,\cdots,B$) for vector $\lambda^T=(\lambda_1^{(s)},\cdots,\lambda_d^{(s)})$ without sampling for each $b=1,\cdots,B$ only a single value from each sample of the marginal posterior distributions. 
In this case \textit{Point} $4$ of Step $2$ in Algorithm \ref{ABCOP} will consist of $S$ matrices $u^{s}$, $s=1,\cdots,S$; consequently, $L_{BEL}$ will provide a set of weights $\omega_{bs}$, $b=1,\cdots,B$ and $s=1,\cdots,S$. 

In this case, the posterior distribution of $\phi$ may be approximated by simply combining the sample from the prior distribution, say $(\phi_1,\cdots,\phi_B)$ with an average of the weights, say $\bar{\omega}_b = \frac{1}{n} \sum\limits_{i=1}^S \omega_{bs}$ for $b=1, \cdots, B$. 
This version of the algorithm will be used in Section \ref{sec:logreturns} as Algorithm \ref{algo:ABCOP-logreturns}. 

This last version of the algorithm is of course more accurate since it considers, at each iteration, the global uncertainty in the marginal distributions . However, its computational burden may be heavy. Algorithm\ref{ABCOP}, therefore, is presented in a more manageable version, where, at each simulation $b=1,\cdots,B$ the marginal posterior distribution are considered as approximated with a sample of size one, randomly selected among the entire marginal posterior sample. 

\textbf{Choice of the priors.}
Prior elicitation is necessary for the marginal parameters $\lambda$ and for the quantity of interest $\phi$. 
While the marginal estimation does not present peculiar issues, 
the elicitation of the prior on $\phi$ could be potentially important. 
However, the most common functional of interest are, in general, defined on a compact space; as a consequence, a default objective choice, in the absence of specific information, is the uniform distribution. Other choices are clearly possible; our simulation studies, not reported here, suggests that the resulting posterior distribution seems to be robust in terms of prior choices. 

The prior on the nonparametric component of the copula is implicitly provided by the use of the exponentially tilted empirical likelihood, as proved in \cite{schen:05}, and discussed in Section \ref{sec:asymptotics}): this aspect, of course, has pros and cons. The main advantage is the ease of elicitation: one need not to elicit about complex aspects of the multivariate dependence structure. This is mainly in the spirit of the so called partially specified models, quite popular in the econometric literature. 
Another obvious advantage is the implied robustness of the method, with respect to different prior opinions about non-essential aspects of the dependence structure. 
The most important disadvantage is its inefficiency when compared to a parametric copula, under the assumption that the parametric copula is the true model. On the other hand, we will see in Section \ref{sec:srho} that the parametric approach may lead to completely wrong results in case of miss-specification. Another aspect to consider is that model selection procedures are not yet fully developed in the copula literature: this is mainly due to the fact that most of the differences among the various copula models refer to the tail behavior, and it is rare to have enough data on the tails to perform reliable model comparison.

\textbf{Existence of an unbiased estimator.}\\
The Bayesian exponentially tilted empirical likelihood is based on moment conditions of the form (\ref{mo-cond}). As already sketched in \S~\ref{3.2}, this kind of conditions implies, at least implicitly, the existence of an unbiased estimator for the quantity of interest. In practical applications, there are often available only asymptotically unbiased estimators. This is the case, for example,  of the Spearman's $\rho$: its sample counterpart $\hat{rho}_n$ is only asymptotically unbiased so the moment condition $E[\hat{\rho}_n-\rho]=0$ is only valid for large samples.

Finally, a note on computational issues: the most demanding step of Algorithm \ref{ABCOP} is the evaluation of $L_{BEL}$. This entails an optimization procedure over the hypercube $[0,1]^{n-1}$, based on Lagrange multipliers, however this may be easily and fastly implemented in \texttt{R} using the generic function \texttt{optim}.

\section{Theoretical Background}
\label{sec:asymptotics}

The method described in the previous section is based on several different theoretical results.

In this section we collect some more theoretical considerations, in order to better clarify advantages and limitations of the proposal.

\textbf{Two-step estimation}

The inferential step has been split into two parts: first, the marginal distributions of the multivariate random variable are estimated; 
then, pseudo-data are created in order to provide a semiparametric estimate of the quantity of interest. 
The ``two-step'' issue is at the core of the often unsatisfactory behaviour of estimation procedures based on the \textit{Inference From the Margins} method, see for a review \cite{joe}, 
Section 10.1: the main drawback of that approach is that it fails to properly account for the uncertainty on the parameter estimates of the marginal distributions. 

However this problem is much less serious in our setting; in fact, we produce, for each coordinate of the multivariate distribution, a sample from the joint posterior distribution of the parameters which appear in that marginal. 
So the actual level of information on those parameters is completely transferred to the \textit{second step} of the procedure, which creates, for each run of the posterior simulation, a different set of pseudo-data and then \textit{takes averages} on them. Provided that the estimation procedure for the marginals is consistent, we are consistently creating multiple ``pseudo-data''.

\textbf{Bayesian semi-parametric interpretation}

In \cite{schen:05} it is argumented and proved that the Bayesian exponentially tilted empirical likelihood has a precise Bayesian interpretation, which we now describe in our context. 
The infinite dimensional parameter space for a copula model can be written as $(C,F_1,\cdots,F_d)$; however, the interest of the analysis is in a low-dimensional function $\phi(C)$. Then the copula $C$ can be represented as $C=(\phi, C^*)$, where $C^*$ belongs to an infinite dimensional metric space $(H,d_H)$ and represents all those aspects of the dependence structure not related to $\phi$. The global nuisance parameter for the model is $\xi=(C^*, F_1, \cdots, F_d)$. 
At each iteration of Algorithm \ref{ABCOP}, i.e. for fixed values of the marginal parameters $\lambda$, the computation of the Bayesian exponentially tilted empirical likelihood may be read as the evaluation of the integrated likelihood of $\phi$, say 
$$L_{BEL}^{(\lambda)}(\phi;u)=\int_{\Xi} L(\phi,\xi;u) d\Pi(\xi),$$
where $u=\{ [u_{ij}]_{ij} , i=1, \cdots ,n, j=1,\cdots,d \}$, $u_{ij}=F_j(x_{ij};\lambda_j)$ and $\Pi(\xi)$ is the nonparametric prior process implicitly induced by the use of the Bayesian exponentially tilted empirical likelihood and specified by Theorem $1$ in \cite{schen:05}. In brief, $\Pi(\xi)$ is a prior process which tends to favour distributions with a high level of entropy. 
Algorithm \ref{ABCOP} takes an average of $L_{BEL}^{(\lambda)}(\phi;u)$ with respect to the posterior distributions of the parameters of the marginals $F_1,\cdots,F_d$. Consequently, it produces an approximation of the integrated likelihood $L(\phi;x)$ which is combined with the genuine prior for $\phi$ in order to obtain $\pi(\phi \mid x)$.

\textbf{Partially specification of the model}

Our model is, in some sense, only partially specified, since we are interested in a specific aspect of the copula.
In order to make Bayes' theorem applicable, we again invoke Theorem $1$ in \cite{schen:05}, which gives a fully Bayesian interpretation of the model. 
It is true however, that the Bayesian exponentially tilted empirical likelihood is a valid approximation of the integrated likelihood for the parameter of interest $\phi$ only when the moment conditions \eqref{mo-cond} are valid. 
It may happen, as already noticed, that available estimators of $\phi$ are only asymptotically unbiased; as a consequence, the moment conditions (and the entire method as well) are valid only for large samples. 
Moreover, the quantities of interest considered in this paper (Spearman's $\rho$ and tail dependence indices $\lambda_L$ and $\lambda_U$) are defined in terms of the copula and the corresponding estimators are based on the empirical copula
\begin{equation}
\label{eq:empcopula}
\widehat{C}_n(u)=\frac{1}{n}\sum_{i=1}^n \prod_{j=1}^d \mathbb{I}_{\left\{\hat{U}_{ij} \leq u_j\right\}}, \qquad  u=(u_1,u_2,\cdots,u_d)\in [0,1]^d
\end{equation}
\noindent where $\hat{U}_{ij}$ are the pseudo-data obtained after the first step of the procedure (the estimation of the marginals). 
It is then clear that, in order to use the Bayesian exponentially tilted empirical likelihood, the empirical copula must be a consistent estimator of the copula. 
In order to check this condition,
suppose we have obtained some estimates of $F_1, \dots , F_d$. For the moment we assume that they come from a Bayesian nonparametric procedure, which is asymptotically equivalent to a procedure based on the empirical marginal cumulative distribution functions.
Pseudo-data are obtained from this first step of the procedure $u_{ij}=\hat{F}_j(x_{ij})$ and they are used for deriving nonparametric estimates of $\phi$; the joint use of a nonparametric procedure for the estimation of the marginals and of an asymptotically unbiased estimator in \eqref{mo-cond} provides a global procedure which is clearly asymptotically equivalent to an estimate based on the empirical copula $\hat{C}_n$.
On the other hand, it is known \citep{fermanian} that $(\hat{C}_n-C)$ is weakly convergent to a Gaussian process in 
$\ell^{\infty}[0,1]$; more precisely, suppose $(X_{11},X_{21},\cdots,X_{d1}), \cdots, (X_{1n},X_{2n},\cdots,X_{dn})$ are independent random vectors with distribution function $F$ and marginals $F_1, F_2,\cdots,F_d$. The empirical estimator of the copula function $C(u_1,u_2,\cdots,u_n)=F(F_1^{-1}(u_1),F_2^{-1}(u_2),\cdots,F_d^{-1}(u_d))$ is 
\begin{equation*}
\hat{C}_n(u_1,u_2,\cdots,u_d)=\hat{F}_n(\hat{F}_{1n}^{-1}(u_1),\hat{F}_{2n}^{-1}(u_2), \cdots, \hat{F}_{dn}^{-1}(u_d)),
\end{equation*}
\noindent where $\hat{F}_n, \hat{F}_{1n}, \hat{F}_{2n},\cdots,\hat{F}_{dn}$ are the joint and marginal empirical distribution functions of the observations. The empirical copula process is defined as 
\begin{equation*}
\mathbb{C}_n=\sqrt{n}(\hat{C}_n-C)
\end{equation*}
and if the $j$-th first order partial derivative exists and is continous on $V_{d,j}=\{ u\in [0,1]^d: 0 < u_j < 1\}$, then $\mathbb{C}_n$ converges  weakly to a Gaussian process \\ $\{\mathbb{G}_C(u_1,u_2,\cdots,u_d), 0<u_1,u_2,\cdots,u_d<1 \}$ in $\ell^{\infty}([0,1]^d)$; for details, see Theorem 3 in \cite{fermanian}. If a Bayesian nonparametric procedure which is asymptotically equivalent to the empirical distribution function is used, we may still advocate Theorem 3 in \cite{fermanian} and the obtained empirical copula is again consistent. 

A similar, but less general, argument may be used if the marginals are estimated parametrically. In this case, the Bayesian procedure will be asymptotically equivalent to a maximum likelihood approach and \cite{joe:05} shows that the two-step procedure based on maximum likelihood estimates is consistent. 

However, for finite sample sizes, there may still be a problem: if we are using a wrong model on the marginals, the entire posterior sample may be misleading and the subsequent step might be biased. This problem is, of course, common to any parametric statistical procedure for copula estimation.  

\section{Monotonic dependence}
\label{sec:srho}

We first illustrate the method in the simple situation $d=2$,  and assuming that the two marginal distributions of the data are known: without loss of generality we assume that they are both uniform in $[0,1]$.
The Spearman's $\rho$ between $X$ and $Y$ is the correlation coefficient among the 
transformed variables $U= F_X(X)$ and $V=F_Y(Y)$ or, in a copula language,
\begin{equation}
\label{rho}
\rho = 12 \int_{0}^1 \int_{0}^1 \big ( C(u,v)- uv \big ) du dv
= 12 \int_{0}^1 \int_{0}^1  C(u,v) du duv - 3.  
\end{equation}
Starting from a sample of size $n$ from a bivariate distribution, say $(x_i, y_i),$ $i=1, \dots,n$, a possible estimator of $\rho$, say $\hat{\rho}_n$, obtained by substituting the empirical copula $\hat{C}_n$ in expression \ref{rho}, is the correlation among 
ranks and it can be written as 
\begin{equation}
\label{rhon}
\hat{\rho}_n = \frac{1}{n} \sum_{i=1}^n \bigg ( \frac{12}{n^2-1} R_i Q_i \bigg ) -3 \frac{n+1}{n-1},
\end{equation}
where 
$$
R_i = \mbox{rank}(x_i) = \sum_{k=1}^n \mathbb{I}(x_k \leq x_i), \quad 
Q_i = \mbox{rank}(y_i) = \sum_{k=1}^n \mathbb{I}(y_k \leq y_i),\quad i=1, \dots, n.$$
We use Algorithm \ref{ABCOP}  to produce a posterior sample for $\rho$, with $\pi(\rho)=\mathcal{U}(-1,1)$: a full description of the application of  Algorithm \ref{ABCOP} in the specific case of the Sperman's $\rho$ in dimension two is described in Algorithm \ref{algo:ABCOP-rho2}; step 1 of Algorithm \ref{ABCOP}  is avoided in this simulated studies, since the simulation are already in the copula space. 

\begin{algorithm}
\caption{ABSCop algorithm - Spearman's $\rho$ }
\label{algo:ABCOP-rho2}

		Given a sample of $n$ pseudo-observations 
		assumed from an unknown copula function 
		
$$
u= \begin{pmatrix}
u_{11} & u_{12} \\
u_{21} & u_{22} \\
\cdots & \cdots \\
u_{n1} & u_{n2} \\
\end{pmatrix} ,
$$

	\begin{algorithmic}[1]
		\For {$b=1,\cdots,B$}
		\State Draw $\rho^{(b)} $ from its prior distribution, for example $\rho^{(b)} \sim \mathcal{U}nif(-1,1)$
		\State Compute a nonparametric estimate of the Spearman's $\rho$:
			$$\hat{\rho}_n = \frac{1}{n} \sum_{i=1}^n \bigg ( \frac{12}{n^2-1} R_i Q_i \bigg ) -3 \frac{n+1}{n-1}$$
			\noindent where $R_i = \sum_{k=1}^n \mathbb{I}(u_{1k} \leq u_{1i}), Q_i  = \sum_{k=1}^n \mathbb{I}(u_{2k} \leq u_{2i}),\quad i=1, \dots, n.$
			\State Compute $L_{BEL}(\rho^{(b)};u)=\omega_b$
			\EndFor
		\State \Return A weighted sample of size $B$ of values for $\rho$, where the weights are defined as the $L_{BEL}$, given the nonparametric estimate $\hat{\rho}_n$. 
		\State Sample with replacement $(\rho^{(b)}\omega_b)$, $b=1,\cdots,B$.
		\end{algorithmic}

\textbf{Output:} a sample of size $B$ of values approximately from the posterior distribution of $\rho$.

\end{algorithm}

The frequentist properties of estimator (\ref{rhon}) have been considered in \cite{borkowf:02}, who shows that the asymptotic variance of $\rho_n$ is
\begin{equation}
\label{eq:sigmarhon}
\sigma^2(\hat{\rho}_n)=144(-9 \psi_1^2+\psi_2+2 \psi_3+2 \psi_4+2 \psi_5),
\end{equation} 
where the $\psi_i$'s are terms linked with the moments of the marginal and joint distributions of $(X_1,Y_1)$ and $(X_2,Y_2)$, two independent random vectors with distribution $F(x,y)$ and marginals $F_1(X)$ and $F_2(Y)$ respectively. In particular
\begin{align*}
& \psi_1=\mathbb{E}[F_1(X_1)F_2(Y_1)], \\
& \psi_2=\mathbb{E}[(1-F_1(X_1))^2(1-F_2(Y_1))^2], \\
& \psi_3=\mathbb{E}[(1-F(X_1,Y_2))(1-F(X_2))(1-F(Y_1))], \\
& \psi_4=\mathbb{E}[(1-F_1(\max\{X_1,X_2\}))(1-F_2(Y_1))(1-F_2(Y_2))], \\
& \psi_5=\mathbb{E}[(1-F_1(X_1))(1-F_1(X_2))(1-F_2(\max\{Y_1,Y_2\}))].
\end{align*}
Consistent estimates of the above quantities are available in \cite{genest3}.

However, in the case of perfect rank agreement, when plugging-in the sample estimates of the  $\psi_j$'s  into expression (\ref{eq:sigmarhon}), one gets a negative number. This phenomenon also occurred in our simulations when data were generated from copulae with a value of $\rho$ close to $1$. 

As an illustration we have simulated $500$ samples of size $n=1000$ from a bivariate Clayton copula with $\rho=0.50$, a Frank copula with $\rho=0.50$, a Gumbel copula with $\rho=0.683$ and a Gaussian copula with $\rho=0.8$. 
For comparative purposes we have also implemented the nonparametric frequentist procedure described in \cite{genest3}, where a confidence interval for the Spearman's $\rho$ is computed based on the asymptotic  distribution of $\rho_n$. 

\begin{figure}
\centerline{
\includegraphics[width=25pc,height=15pc]{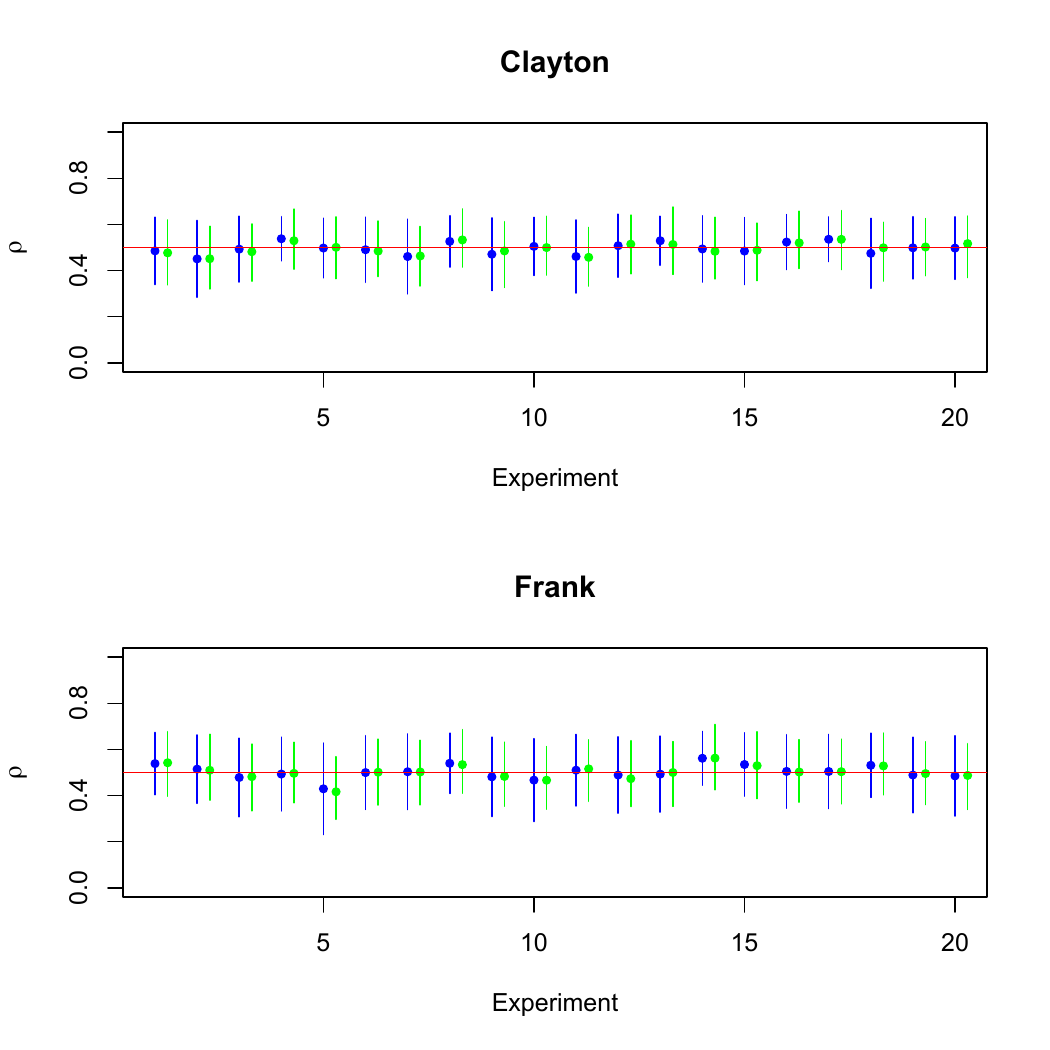}
}
\caption{Comparison between frequentist (blue) and Bayesian estimates (green). $20$ out of $500$ experiments with simulations from a Clayton copula (above) and a Frank copula (below) ($n=1000$); the true value of $\rho$ is $0. 5$ in both cases (red lines), the circles represent the point estimates and the lines represent the (confidence or credible) intervals. 
}
\label{fig:clayfrank_k2}
\end{figure}

\begin{figure}[h]
\centerline{
\includegraphics[width=28pc,height=15pc]{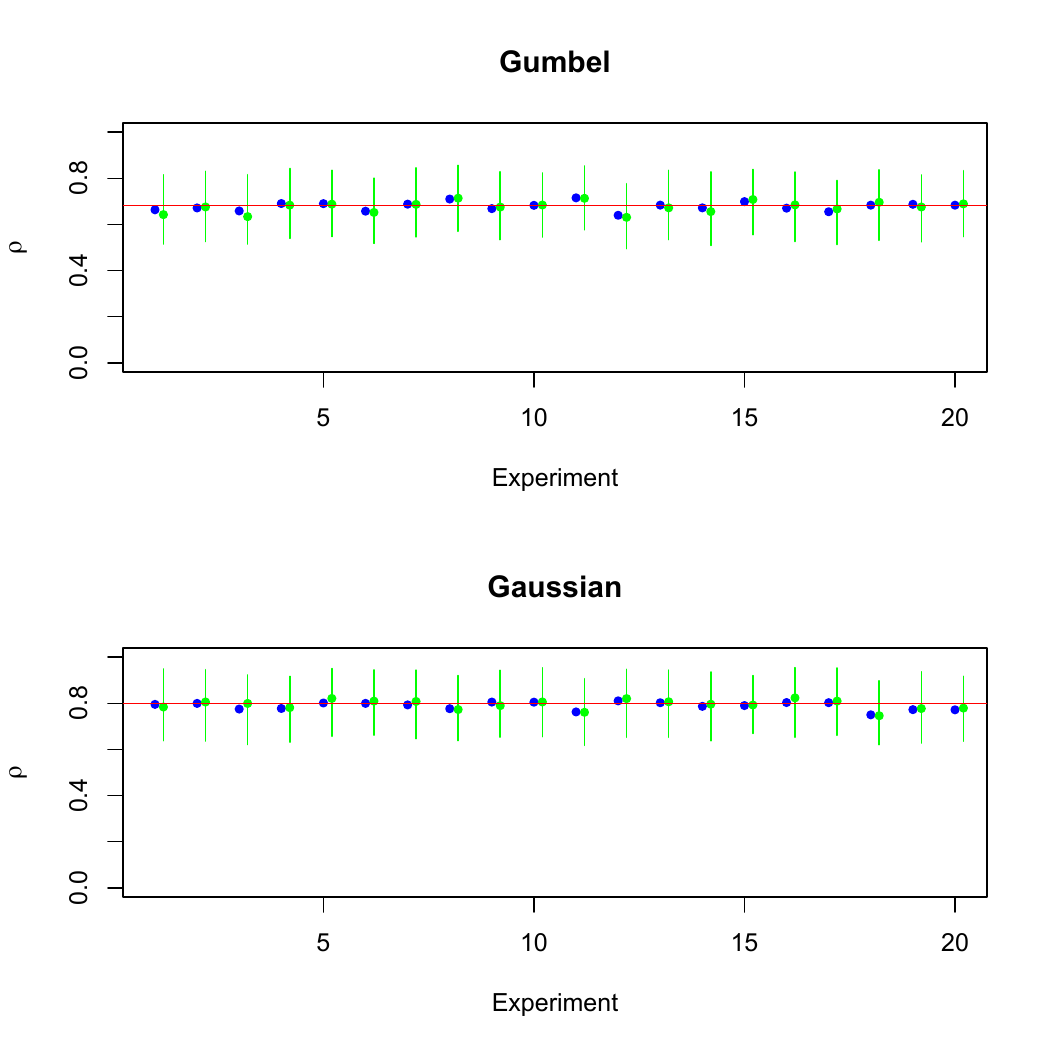}
}
\caption{Comparison between frequentist (blue) and Bayesian estimates (green). $20$ out of $500$ experiments with simulations from a Gumbel copula with $\rho=0.68$ (above) and a Gaussian copula with $\rho=0.8$ (below) ($n=1000$); the true values are represented by the red horizontal lines, the circles represent the point estimates and the lines represent the (confidence or credible) intervals. The frequentist intervals are impossible to be computed because the variance is estimated by a negative value in both cases. 
}
\label{fig:gumbgauss_k2}
\end{figure}

Figure \ref{fig:clayfrank_k2} compares the frequentist behaviour of the confidence procedure and our proposal, in the case of Clayton and Frank copulae, Figure \ref{fig:gumbgauss_k2} presents the same comparison for Gumbel and Gaussian copulae in the case of stronger positive dependence. One can notice that, for large values of $\rho$ (i.e. close to 1), the frequentist estimate of the variance is negative in most cases. As a consequence, confidence intervals can not be produced. 

Figure \ref{fig:clayfrank_k2} and Table $\ref{tab:ICk2}$ show that our method produces shorter interval estimates of $\rho$, compared to the frequentist approach, while maintaining the correct coverage. Notice that in this simulation study, we have used uniform marginals; this implies that this improvement is only due to the different way of dealing with the dependence structure and not on the accounting for uncertainty in the marginal estimation. The posterior median is always very close to the empirical value $\hat{\rho}_n$, however Table \ref{tab:ICk2} shows that the average length of the frequentist intervals is larger than the corresponding Bayesian credible intervals when the frequentist procedure is valid, i.e. the estimated variance is non-negative.

As the true value of $\rho$ increases, the frequentist estimate of the variance tends to be negative ($98. 4\%$ of the experiments for the Gumbel copula with $\rho=0. 68$ and $100\%$ of the experiments for the Gaussian copula with $\rho=0. 80$); on the other hand, our procedure performs equally well. 
The proportion of frequentist intervals with larger length than the corresponding Bayesian interval is $0.564$ for the Clayton copula (with $\rho=0.5$) and $0.892$ for the Frank copula (with $\rho=0.5$); the coverage in the other two cases cannot be evaluated because of the negative frequentist estimate of the variance.

\begin{table}[]
\centering
\caption{Simulations from different copulae: average length and empirical coverage of the intervals obtained both via frequentist and Bayesian methods, based on $500$ repetitions of the experiment}
\label{tab:ICk2}
\begin{tabular}{cc|cc}
                                &                 & \textbf{Ave. Length} & \textbf{Coverage} \\  \hline
\textbf{Clayton ($\rho=0.50$)}   & \textit{Freq.}  & $0. 2664$              & $0. 998$             \\
\textbf{}                       & \textit{Bayes.} & $0. 2597$              & $1. 000 $            \\ \hline
\textbf{Frank ($\rho=0. 50$)}     & \textit{Freq.}  & $0. 3172 $             & $1. 000$             \\
\textbf{}                       & \textit{Bayes.} & $0. 2735 $              & $1. 000 $            \\ \hline
\textbf{Gumbel ($\rho=0. 68$)}   & \textit{Freq.}  & -              & -             \\
\textbf{}                       & \textit{Bayes.} & $0. 2966$               & $1. 000 $            \\ \hline
\textbf{Gaussian ($\rho=0. 80$)} & \textit{Freq.}  & -               & -             \\
                                & \textit{Bayes.} & $0. 2931$               & $1. 000$            
\end{tabular}
\end{table}

Another advantage of using our semiparametric approach is its robustness with respect to model miss-specification. To show this, we have compared our results with a fully parametric approach based on standard MCMC algorithms. In particular, we have re-used the previously simulated data under the following assumptions:

\begin{itemize}
\item Clayton copula and $\theta \sim \mathcal{TN}(0,10,-1,\infty)$;
\item Gumbel copula and $\theta \sim \mathcal{TN}(1,10,1,\infty)$;
\item Frank copula and $\theta \sim \mathcal{N}(0,10)$;
\end{itemize}

\noindent where $\mathcal{TN}(\mu,\sigma,a,b)$ is a truncated normal distribution with mean $\mu$, standard deviation $\sigma$ and truncation in $[a,b]$. 
Finally, the approximated posterior distributions for the copula parameters are transformed in the corresponding posterior distributions for the Spearman's $\rho$ relative to that particular copula. 

\begin{figure}
\centerline{
\includegraphics[width=25pc,height=15pc]{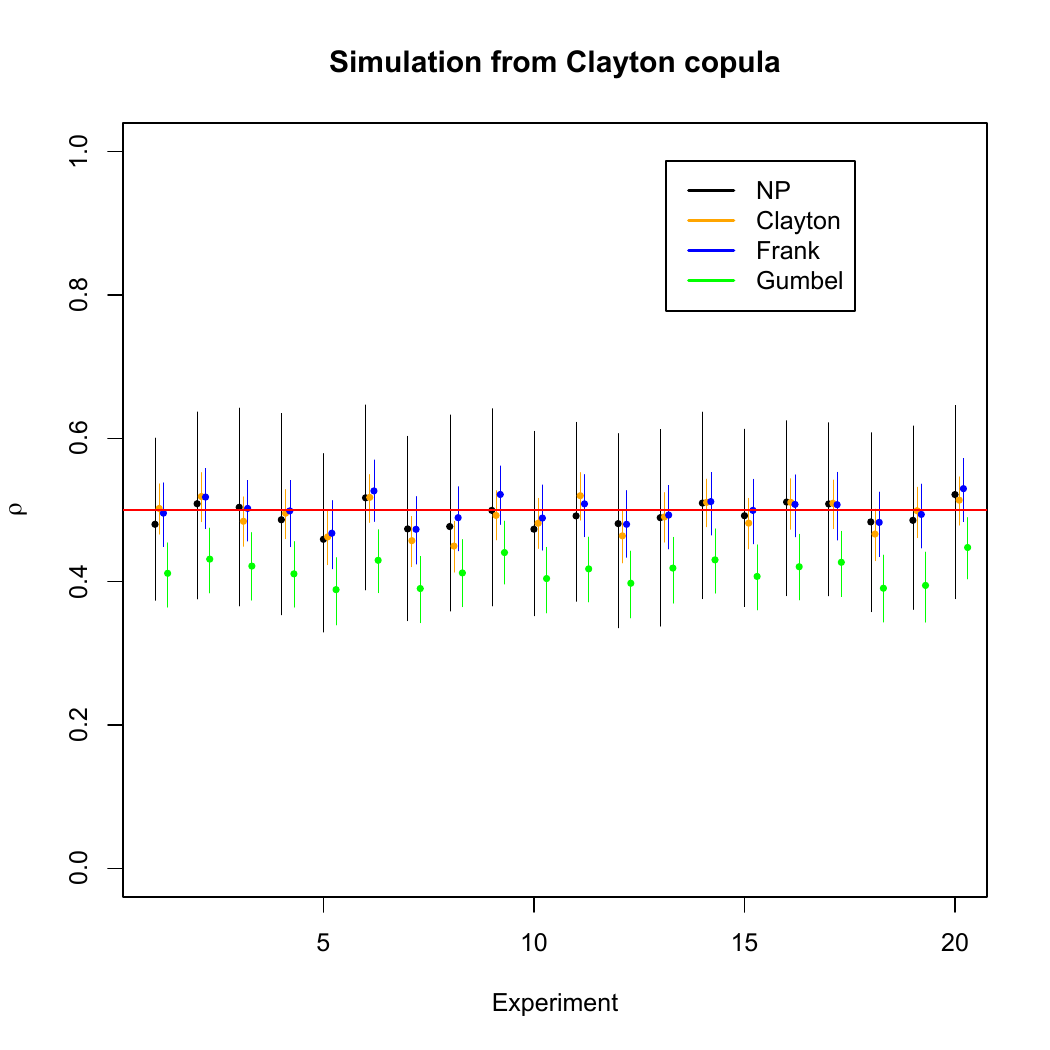}
}
\caption{Bayesian point estimates (points) and credible intervals for $20$ out of $500$ experiments with data from a Clayton copula with $\theta=1.076$, obtained by specifying a Clayton model (orange), a Frank model (blue) and a Gumbel model (green) or by using our semiparametric approach (black). The solid red line represent the true value. 
}
\label{fig:np_param_clayton_k2}
\end{figure}

\begin{figure}[h]
\centerline{
\includegraphics[width=28pc,height=15pc]{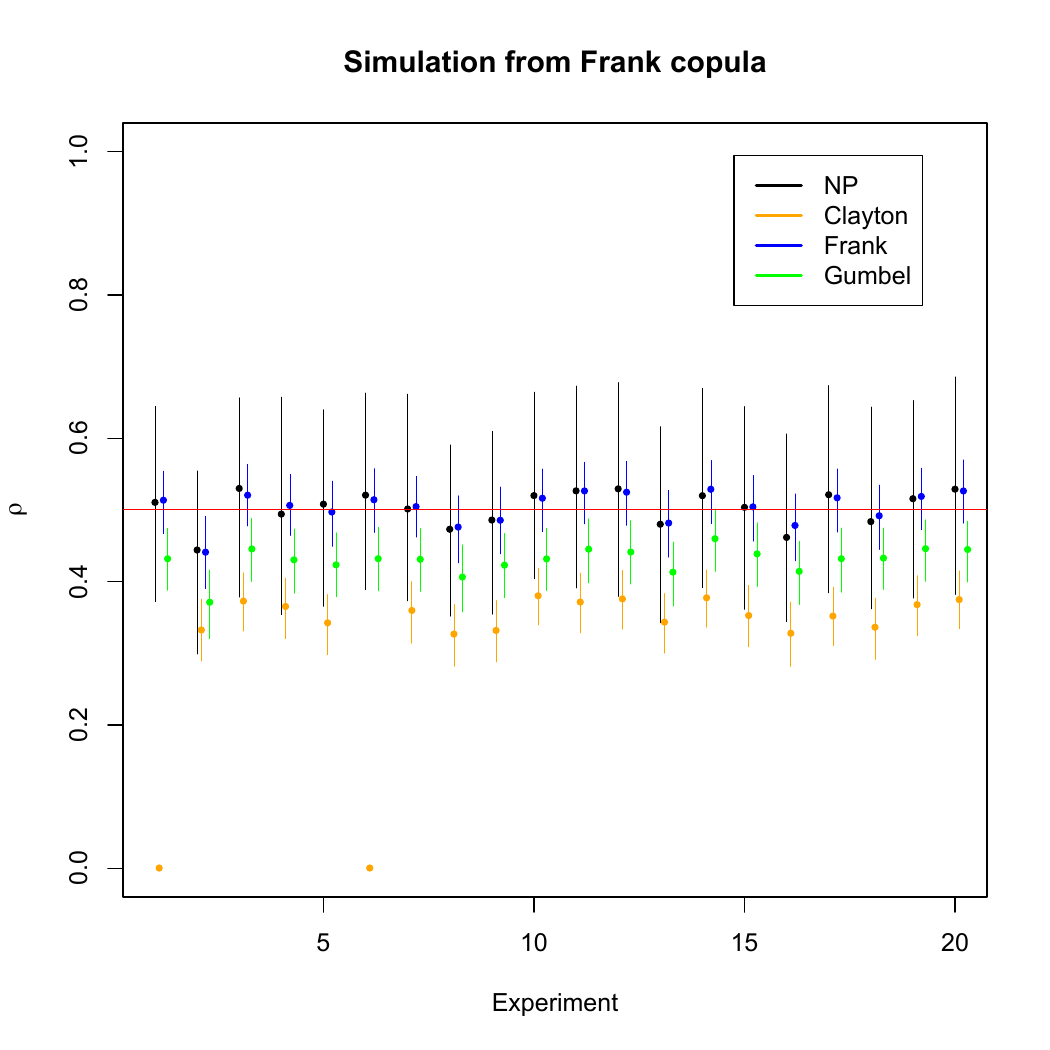}
}
\caption{As in Figure \ref{fig:np_param_clayton_k2}, with simulation from a Frank copula with $\theta=3.45$.
}
\label{fig:np_param_frank_k2}
\end{figure}

\begin{figure}[h]
\centerline{
\includegraphics[width=28pc,height=15pc]{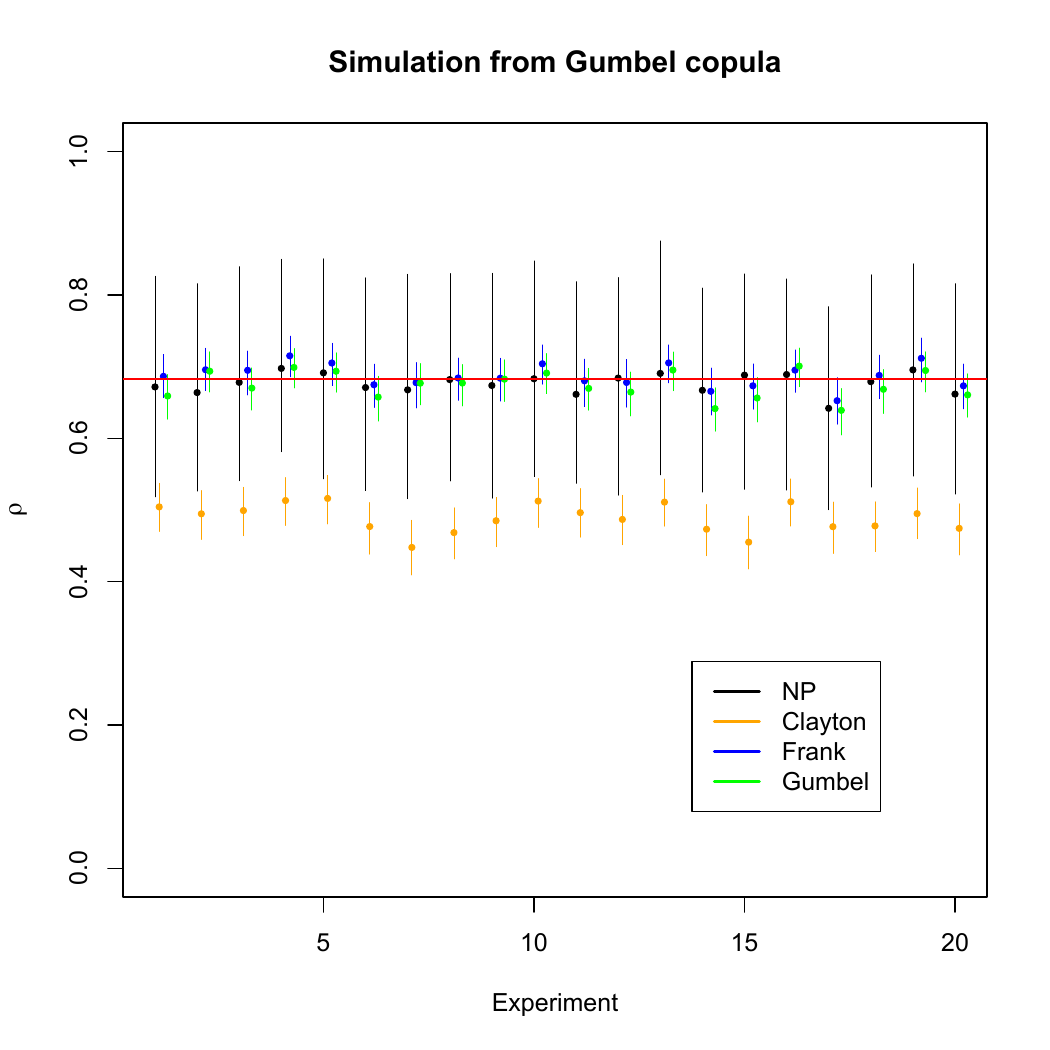}
}
\caption{As in Figure \ref{fig:np_param_clayton_k2}, with simulation from a Gumbel copula with $\theta=2$.
}
\label{fig:np_param_gumbel_k2}
\end{figure}

Figures \ref{fig:np_param_clayton_k2}, \ref{fig:np_param_frank_k2}, \ref{fig:np_param_gumbel_k2} and Table \ref{tab:np_param_k2} show the results of the simulations. It is evident that, although a parametric model produces shorter credible intervals, the choice of the particular parametric copula is crucial. The semiparametric method is clearly the most robust. 

\begin{table}[]
\centering
\caption{Simulations from different copulae: average coverage of the $0.95$ equal tails Bayesian credible intervals obtained in $500$ repetitions of the experiment.}
\vspace{0.1in}
\label{tab:np_param_k2}
\begin{tabular}{c|ccc}
           & \textbf{True Clayton} & \textbf{True Frank} &\textbf{True Gumbel} \\  
\textbf{Assumption}	& $\theta=1.076$ & $\theta=3.45$ & $\theta=2$ \\ \hline
\textit{Clayton}    & $0.852$ & $0.000$ & $0.000$ \\
 \hline
\textit{Frank}       & $0.920$ & $0.938$ & $0.838$ \\
 \hline
\textit{Gumbel}    & $0.052$ & $0.082$ & $0.878$ \\
 \hline
\textit{ABSCop}    & $0.999$ & $0.999$ & $0.999$                 
                                         
\end{tabular}
\end{table}

\section{Tail Dependence}
\label{sec:TD}

Multivariate dependence may be a complicated object. Popular measures like the Spearman's $\rho$ or the Kendall's $\tau$ can only capture some aspects of it. For example, dependencies between extreme negative stock returns or large portfolio losses are better explained by tail dependence indices  \citep{sibuya:60}. Several studies show that, in particular in volatile markets, tail dependence is a useful tool to study the behaviour of extremal data in finance. See, for example, \cite{ane:kharoubi:03}.
Unfortunately the tail dependence is delicate to estimate, mostly because of the limited amount of available data in the tails of the distribution. 

The concept of tail dependence describes the idea of concordance in the tails of the bivariate distribution, i.e. the amount of dependence in the lower-left quadrant tail or upper-right quadrant tail. 
The upper and lower tail dependence indices are defined in terms of the survival function:

\begin{align}
\label{eq:tdc}
\lambda_U & = \lim_{v \rightarrow 1^{-}} \Pr \left\{  F_X (X) > v | F_Y (Y) > v \right\}, \nonumber \\
\lambda_L & = \lim_{v \rightarrow 0^{+}} \Pr \left\{  F_X (X) \leq v | F_Y (Y)  \leq v \right\},
\end{align}

\noindent provided the limits exist. $(X,Y)$ are said to be upper tail dependent if $\lambda_U > 0$ and upper tail independent if $\lambda_U > 0$. Similar definitions apply for $\lambda_L$.
These definition clarify the concept of tail concordance: the upper (lower) tail dependence index is close to one if the probability that the marginal distribution of one variable exceeds a high (low) threshold given that the marginal distribution of the other variable exceeds a high (low) threshold is close to one. 

However, the tail dependence indices, as defined in \eqref{eq:tdc}, only depend on the copula structure:

\begin{equation}
\label{tail-d}
\lambda_U=\lim_{v \rightarrow 1} \frac{1-2v+C(v,v)}{1-v}, \qquad 
\lambda_L=\lim_{v\rightarrow 0} \frac{C(v,v)}{v} .
\end{equation}

\noindent and, therefore, they may be estimated by using the Bayesian approach proposed in Section \ref{sec:proposal}.

It is necessary to choose a nonparametric estimator of $\lambda_U$ and $\lambda_L$ in order to apply Algorithm \ref{ABCOP}. 
For a review on the parametric and nonparametric estimation of the tail dependence indices, see \cite{frahm}. 
Among the many proposals, here  we consider, as a benchmark, the estimator given in \cite{frahm} as a special case of the one proposed in \cite{joe:smith:weissman:92}:

$$\hat{\lambda}_L  =\frac { \hat{C}_n\left(\frac{k}{n},\frac{k}{n}\right) }{\frac{k}{n}}, 
\hat{\lambda}_U  =2- \frac { 1-\hat{C}_n\left(\frac{n-k}{n},\frac{n-k}{n}\right) }{1-\frac{n-k}{n}}, \quad
,
$$
\noindent where $\hat{C}_n$ is the empirical copula, and $0< k \leq n$ is a parameter tuned by the experimenter. A typical choice, motivated in \cite{joe:smith:weissman:92}, is $k=\sqrt{n}$. 
\cite{schmidt:stadtmuller:06} prove strong consistency and asymptotic normality for these estimators: the moment conditions for the application of the empirical likelihood approach are, therefore, (asymptotically) valid. 

\cite{schmidt:stadtmuller:06} have also derived the asymptotic variance of $\hat{\lambda}_L$ and $\hat{\lambda}_U$. However, these expressions are of limited use since they depend on unknown quantities. To circumvent this problem, they propose to use the variance of the tail dependence coefficient of a copula for which the same quantities are easy to compute. Nevertheless this method does not provide any quantification of the potential error, which is essential in the particular case of tail dependence coefficients, for which the estimation procedure is, in general, based on a small proportion of the available data. 
In contrast, with our approach, we are able to provide an approximation of the entire posterior distribution of the index, which can then be summarized in different ways.

\begin{figure}
\centerline{
\includegraphics[width=25pc,height=15pc]{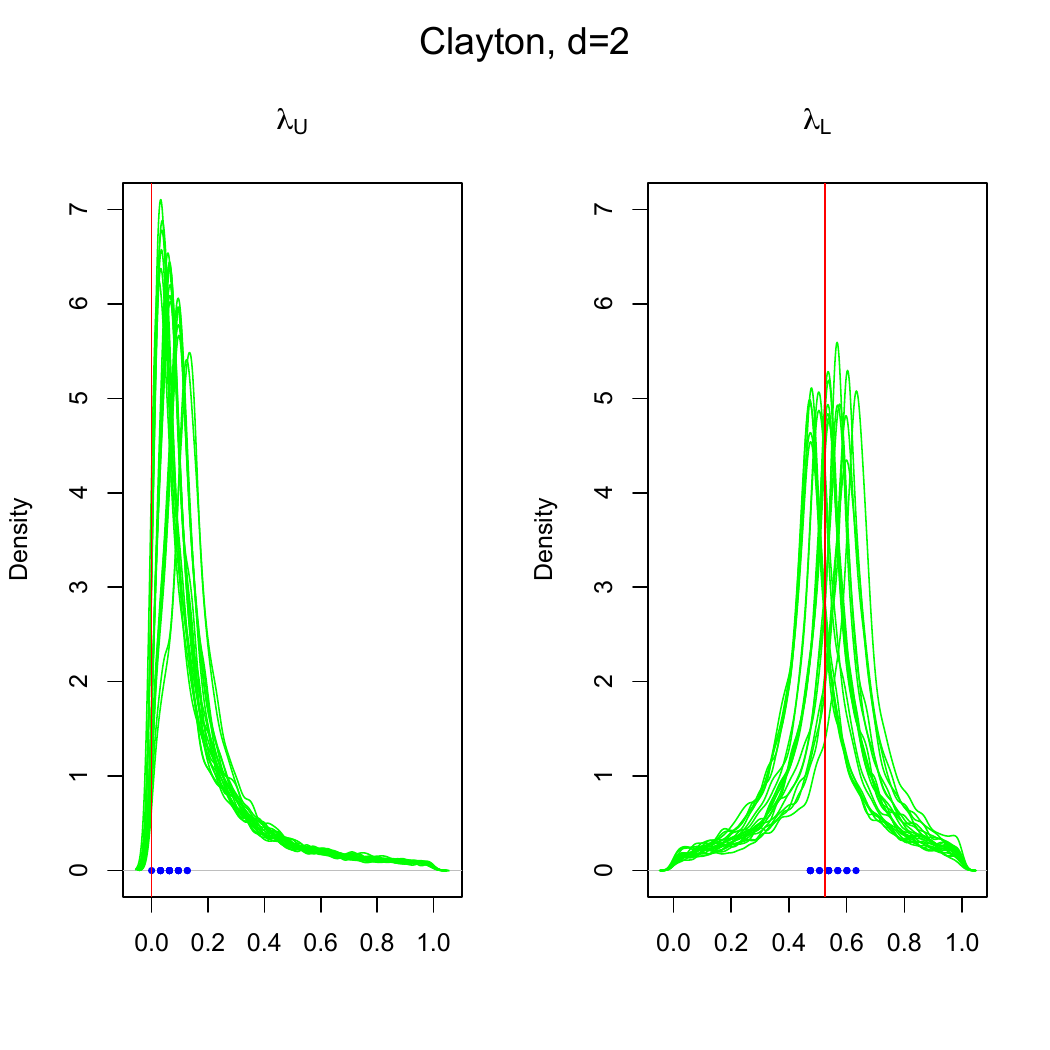}
}
\caption{Comparison between frequentist (blue) and Bayesian (green) estimates for $\lambda_U$ (left) and $\lambda_L$ (right). $20$ out of $500$ simulations from a Clayton copula with $\theta=1.076$ ($n=1000$); the circles represent the frequentist point estimates, the lines represent the approximated posterior distributions. The true values are $\lambda_U^{true}=0$ and  $\lambda_L^{true}=2^{-\frac{1}{\theta}}$ (red lines).}
\label{fig:lambda_clayton}
\end{figure}

Figure \ref{fig:lambda_clayton} shows the approximated intervals for the frequentist (obtained via a bootstrap estimation of the variance) and the Bayesian procedure for simulations from a Clayton copula with $\theta=1.076$ ($\lambda_{L}=2^{-1/\theta}=0.525$ and $\lambda_U=0$). Bayesian intervals are always wider than the corresponding frequentist ones. Nevertheless, the coverage of the frequentist intervals is, on average, around $0. 10$, far from the nominal $0. 95$, which is reached by the Bayesian estimates. See also the Supplementary Material for other examples.   

\section{Multivariate Analysis}
\label{sec:multianalysis}

The extension of the proposed procedure to the multivariate case is straightforward, and no further theoretical issues arise. On the other hand, a broadly satisfactory solution in the frequentist approach has not yet been fully developed. 

It is important to notice that the way to describe the multivariate dependence with low-dimensional measures is still an open problem, since the number of combinations among variables increases with the dimension; as a consequence, there are several ways to define a multivariate measure of dependence. This partially explains why multivariate functionals of dependence are less used, in practice, than their bivariate counterparts.

Formula \eqref{rho} provides one of the possible ways to express the Spearman's $\rho$ and it suggests to interpret
it as a measure of expected distance between the actual copula and the \textit{independence} copula $\Pi(u_1,\cdots,u_d)=u_1 \times \cdots \times u_d$. In this sense, the extension to the $d$-dimensional setting is: 

\begin{equation}\label{rho1}
\rho_1 =\frac{
\int_{[0,1]^d} \left( C({u}) - \Pi({u}) \right )d {u}} {\int_{[0,1]^d} \left( M({u}) - \Pi({u}) \right )d {u}} 
	= h(d) \left\{ 2^d \int_{[0,1]^d}C({u})d {u}-1 \right\},
\end{equation}

\noindent where $M({u})=\min(u_1, u_2, \dots, u_d)$ is the upper Fr\'echet- Hoeffding bound, and $h(d) = (d+1)/ \{2^d-(d+1)\}$.
Other definitions of the Spearman's $\rho$ exist in the literature \citep{schmid:schmidt:07}, for instance:
\begin{equation}
\label{rho2}
\rho_2= h(d) \left\{ 2^d \int_{[0,1]^d} \Pi({u}) dC({u}) -1 \right\}. 
\end{equation}

Finally, a third generalization 
of $\rho$ can be obtained as the average of all the bivariate $\rho$'s.
This expression appears in \cite{joe90}; its rationale is different from those of (\ref{rho1}) and 
(\ref{rho2}), and we will not consider it. 
If $d=2$, then $\rho_1=\rho_2$, but this relation does not necessarily hold in general.  

Nonparametric estimators of the multivariate $\rho_k$ for $k=1,2$ to be used in Algorithm \ref{ABCOP} are again based on the use of the empirical copula \eqref{eq:empcopula} in expressions \eqref{rho1} and \eqref{rho2}: 

\begin{align*}
& \hat{\rho}_{1n}=h(d)\left\{ 2^d \int_{[0,1]^d} \hat{C}_n( {u})d {u} -1\right\}=h(d)\left\{\frac{2^d}{n}\sum_{i=1}^n \prod_{j=1}^d (1-\hat{U}_{ij}) -1 \right\}, \\
& \hat{\rho}_{2n}=h(d) \left\{ 2^d \int_{[0,1]^d} \Pi( {u}) d \hat{C}_n ( {u}) -1 \right\} = h(d) \left\{ \frac{2^d}{n} \sum_{i=1}^n \prod_{j=1}^d \hat{U}_{ij} -1 \right\} .
\end{align*}
Asymptotic properties of these estimators are explored and assessed in \cite{schmid:schmidt:07}. In particular 
it is known that 
\begin{equation*}
\sqrt{n}(\hat{\rho}_{kn}-\rho_k) \overset{\cdot}{\sim} \mathcal{N}(0,\sigma_k^2), \qquad k=1,2.
\end{equation*}

\noindent The expressions for $\sigma_k^2$, $k=1,2$ are given in \cite{schmid:schmidt:07}. 
The variances of the above estimators can be analytically computed only in very few cases. In general, they depend on unknown quantities which must be estimated, for example via bootstrap methods. Bootstrap estimators of $\rho_1$ and $\rho_2$ have been proved to be consistent \citep{schmid:schmidt:06}: on the other hand, the bootstrap estimators of the variances tend to dramatically underestimate the variability of $\hat{\rho}_{kn}$, $k=1,2$. 
We have performed several simulation experiments and our results always indicate that the coverage of the resulting confidence intervals for both $\rho_1$ and $\rho_2$ may be quite far from the nominal value and that the severity of the problem typically depends on the specific copula we sampled from. 
On our approximate Bayesian side, once an estimator of the multivariate version of $\rho$ is available, it is easy to apply the procedure presented in  Section \ref{sec:proposal}, with no particular modifications.

\begin{figure}
\centerline{
\includegraphics[width=25pc,height=15pc]{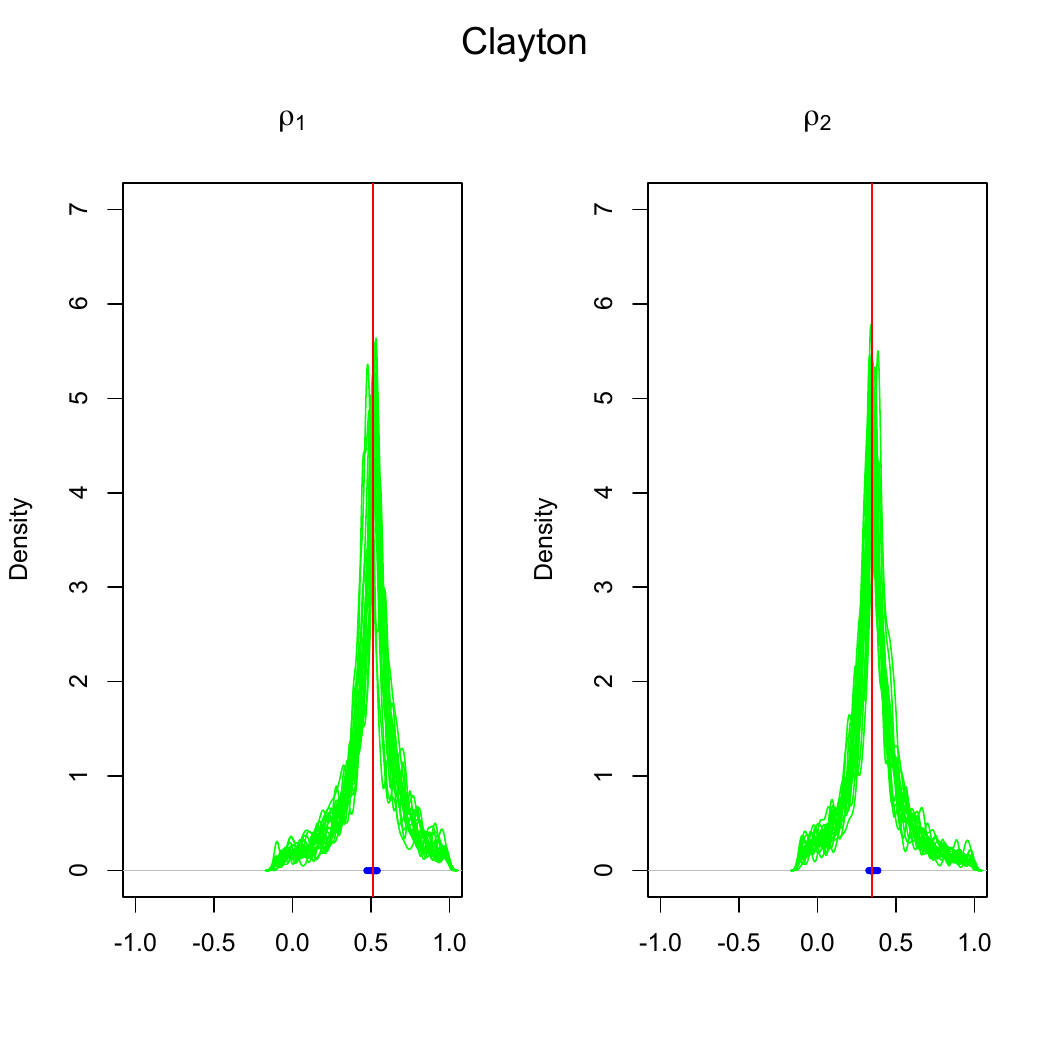}
}
\caption{Comparison between frequentist (blue) and Bayesian (green) estimates of $\rho_1$ (left) and $\rho_2$ (right) as defined in equations \eqref{rho1} and \eqref{rho2}. $20$ out of $500$ experiments with simulation from a Clayton copula with $\theta=1.076$ ($n=1000$); the true values are the red vertical lines, the blue points represent the frequentist point estimates and the green lines represent the approximated posterior distributions. }
\label{fig:rho6_clayton}
\end{figure}

Figure \ref{fig:rho6_clayton} shows the results of a simulation study with a Clayton copula with $\rho_1=0.514$ and $\rho_2=0.346$.
Frequentist intervals obtained via a bootstrap estimate of the variance of $\hat{\rho_k}$, $k=1,2$ 
are always very narrow; the estimated coverage is about $5.8\%$ and it tends to further decrease as the degree of the dependence increases. 
It must be said that, at least for reasonably large sample sizes, the frequentist point estimates of $\rho_1$ and $\rho_2$ are always very precise: however, the methods for evaluating their standard errors seem to be seriously biased downward. 
We discuss  examples of other copula families in the Supplementary Material, where the coverage can be even worse than in the case studied here. 

We notice that the average length of the confidence intervals for ${\rho_1}$ and ${\rho_2}$ does not change significantly as the dimension $d$ of the data increases.
Table \ref{tab:multirho_incrd} shows the average length of the estimated confidence intervals for ${\rho_1}$ and ${\rho_2}$ and the average length of the corresponding (approximated) Bayesian equal tailed $95\%$ credible intervals. One can notice that the average length of the Bayesian intervals shows a decreasing pattern as $d$ increases.
Our conjecture is that, for fixed $n$, the amount of information on a scalar quantity of interest tends to increase with the dimension of the data; the same phenomenon is less significant on the frequentist side because the length of intervals is always very small.

\begin{table}[]
\centering
\caption{Average lengths of the confidence intervals (based on a bootstrap estimator of the variance of the estimates) and of the corresponding Bayesian $95\%$ credible intervals obtained in $50$ repetitions of each experiment of dimension $d$ by simulating data from a Clayton copula with $\theta=1.076$.}
\label{tab:multirho_incrd}
\begin{tabular}{c|cccc}
                & \textbf{$\hat{\rho_1}^{freq}$} & \textbf{$\hat{\rho_2}^{freq}$} & \textbf{$\hat{\rho_1}^{Bayes}$} & \textbf{$\hat{\rho_2}^{Bayes}$} \\ \hline
\textbf{$d=2$}  & 0.0032                         & 0.0032                         & 1.1933                          & 1.1801                          \\ 
\textbf{$d=3$}  & 0.0026                         & 0.0026                        & 1.0844                          & 1.0853                          \\ 
\textbf{$d=4$}  & 0.0026                         & 0.0026                         & 0.9495                          & 0.9594                          \\ 
\textbf{$d=5$}  & 0.0027                         & 0.0027                         & 0.8728                          & 0.8914                          \\ 
\textbf{$d=6$}  & 0.0027                         & 0.0027                         & 0.8211                          & 0.8224                          \\ 
\textbf{$d=7$}  & 0.0030                         & 0.0030                         & 0.8022                          & 0.7882                          \\ 
\textbf{$d=8$}  & 0.0031                         & 0.0031                         & 0.7828                          & 0.7541                          \\ 
\textbf{$d=9$}  & 0.0032                         & 0.0032                         & 0.7680                          & 0.7492                          \\ 
\textbf{$d=10$} & 0.0035                         & 0.0035                         & 0.7558                          & 0.7439                          \\ 
\textbf{$d=25$} & 0.0047                         & 0.0047                         & 0.7462                          & 0.7480                          \\ 
\textbf{$d=50$} & 0.0073                         & 0.0073                         & 0.7299                          & 0.7634                         
\end{tabular}
\end{table}

Multivariate extensions of tail dependence indices are not yet fully developed. An interesting proposal for Archimedean copulae is discussed in  \cite{diberna}; consider a random vector $X=(X_1,X_2,\cdots,X_d)$, more precisely its version lying in the copula space $U=(U_1,U_2,\cdots,U_d)$ and denote by $I$ the set $\{1,2, \dots,d\}$. Consider two non-empty subset of $I$, $I_h \subset I$ and $\bar{I}_h = I \setminus I_h$ of cardinality $h \geq 1$ and $d-h \geq 1$. Provided that the limits exist, the multivariate tail dependence coefficients are given by

\begin{align*}
\lambda_U^{I_h, \bar{I}_h}&=\lim_{u\rightarrow 1^{-}} \Pr\{U_i \geq u , i\in I_h | U_i \geq u , i\in \bar{I}_h \} \\
\lambda_L^{I_h, \bar{I}_h}&=\lim_{u\rightarrow 0^{+}} \Pr\{U_i \leq u , i\in I_h | U_i \leq u , i\in \bar{I}_h \}  \}
\end{align*}

\noindent which describe the relative deviation of upper or lower tail probabilities of a random vector from similar tail probabilities of a subset its component. These coefficients are not uniquely defined, except that in the case of dimension $d=2$, since they depend on the choice of subsets $I_h$ and $\bar{I}_h$ (in the case $d=2$, $h=d-h=1$ necessarily. 
\cite{deluca:rivieccio:12} makes a particular choice of the subsets $I_h$ and $\bar{I}_h$ and define the tail dependence coefficients as:

\begin{align*}
\lambda_U &=\lim_{u\rightarrow 1^{-}} \Pr\{F_1(X_1) \geq u | F_2(X_2) \geq u , \cdots, F_d(X_d) \geq u    \} \\
\lambda_L &=\lim_{u\rightarrow 0^{+}} \Pr\{F_1(X_1) \leq u | F_2(X_2) \leq u , \cdots, F_d(X_d) \leq u    \}
\end{align*}

\noindent by allowing for a simple copula representation

\begin{align*}
\lambda_U &=\lim_{u\rightarrow 1^{-}} \frac{C(1-u,\cdots,1-u)}{(1-u)} \\
\lambda_L &=\lim_{u\rightarrow 0^{+}} \frac{C(u,\cdots,u)}{(u)}
\end{align*}

\noindent which is also used in \cite{salazar:ng:2015}. These definitions will be used throughout the paper. 

We have applied Algorithm \ref{ABCOP} to the particular problem of nonparametrically estimating the tail dependence indices for data simulated from a Clayton, a Frank and a Gumbel copula. We have also derived the analytical formulas for $\lambda_U$ and $\lambda_L$ for these copulae, which are available in Table \ref{tab:multi_taildep}. 

As an alternative, \cite{diberna} propose to estimate the multivariate tail dependence indices through estimation of the copula generator; however, if we assume to have no information about the shape of the copula function, it is difficult to assess the estimation error in this way. 
On the other hand, our approach may be easily extended to this multivariate setting.  

\begin{table}[]
\centering
\caption{Tail dependence indices in multivariate Archimedean copulas, following the general definition in \cite{salazar:ng:2015}.}
\label{tab:multi_taildep}
\smallskip
\begin{tabular}{c|c|c|c}
\textbf{Copula} & $C(u_1,\cdots,u_d)$ & $\lambda_L$     & $\lambda_U$                                                                                   \\ 
\hline
& & & \\
Clayton         & $(1 + \theta \left[\frac{1}{\theta} \sum_{j=1}^d (u_j^{-\theta}-1) \right])^{-1/\theta}$                            & $d^{-1/\theta}$ & 0                                                                                                      \\
& & & \\
Frank           & $-\frac{1}{\theta}\log\left[ 1 + \frac{\prod_{j=1}^d (\exp(-\theta u_j)-1)}{[\exp(-\theta)-1]^{d-1}}\right]$                            & 0                        &  0 \\ 
& & & \\
Gumbel          & $\exp \left[- \left[\sum_{j=1}^d (-\log u_d)^\theta \right] ^{1/\theta}\right]$                           & 0                        & $ \sum_{r=1}^{d} (-1)^{r+1} { {d}\choose{r} } r^{1/\theta}$                                                                                                      \end{tabular}
\end{table}

\begin{figure}
\centerline{
\includegraphics[width=28pc,height=15pc]{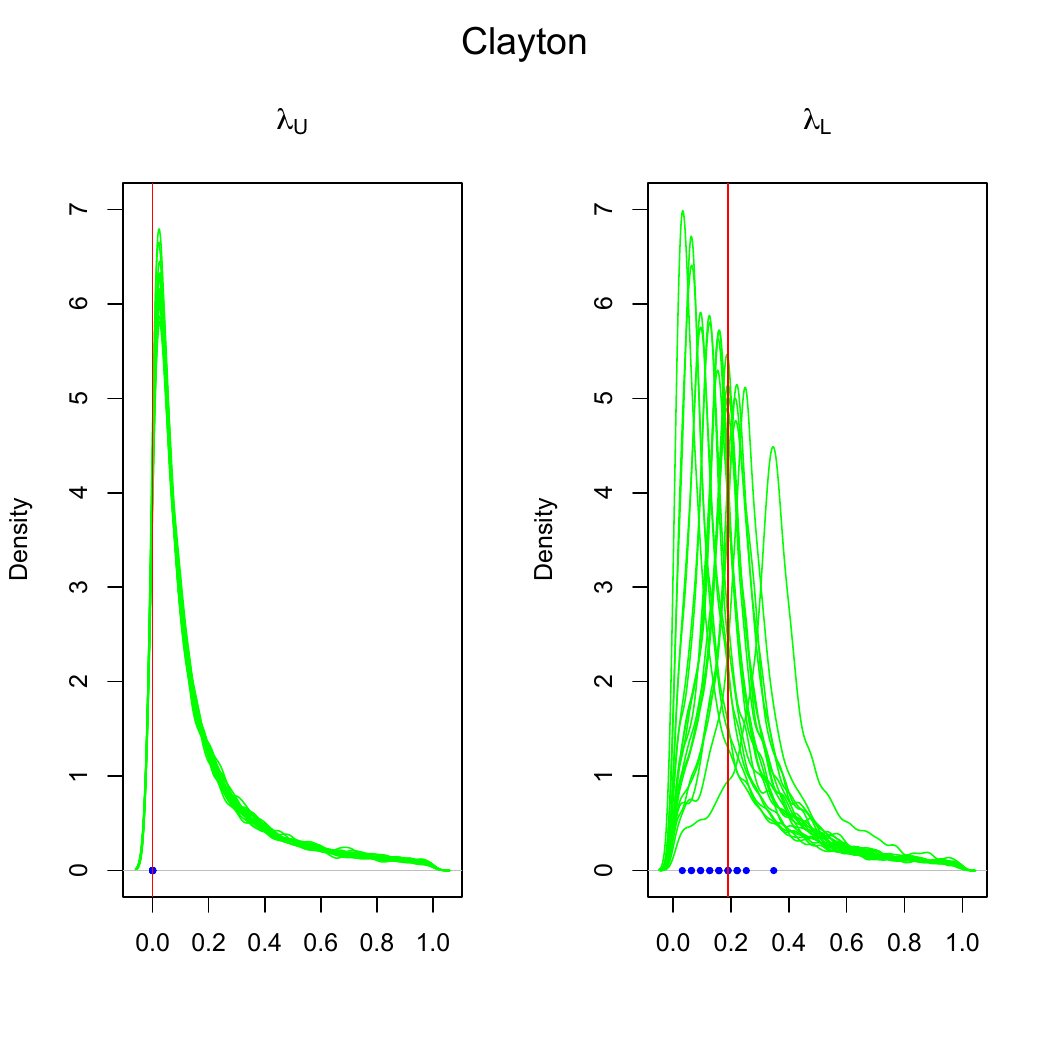}
}
\caption{Approximated posterior distributions (green lines) of $\lambda_U$ (left) and $\lambda_L$ (right) for simulations from a six-dimension Clayton copula with $\theta=1.076$ in $20$ out $500$ experiments. The blue points are the frequentist point estimates, the red lines are the true values, $\lambda_U^{true}=0$ and $\lambda_L^{true}=6^{-1/\theta}$.}
\label{fig:multiTD_clayton}
\end{figure}

Figure \ref{fig:multiTD_clayton} shows the approximated posterior distributions for $\lambda_U$ and $\lambda_L$ obtained with Algorithm \ref{ABCOP} for $20$ out of $500$ experiments with simulations from a Clayton copula with $\theta=1.076$. While the frequentist procedure seems very precise in the case of no tail dependence, there is more variability in the estimates when there is tail dependence. In particular, the lack of realiable methods of evaluating the uncertainty linked to the estimates is a crucial problem: in $500$ repetitions of the experiment the range of variation of the point estimates for $\lambda_L$ is $[0.000,0.379]$. 

Simulations from other types of copulas are available in the Supplementary Material. 

\section{Example: dependence among financial log-returns}
\label{sec:logreturns}

\begin{figure}
\centerline{
\includegraphics[width=28pc,height=15pc]{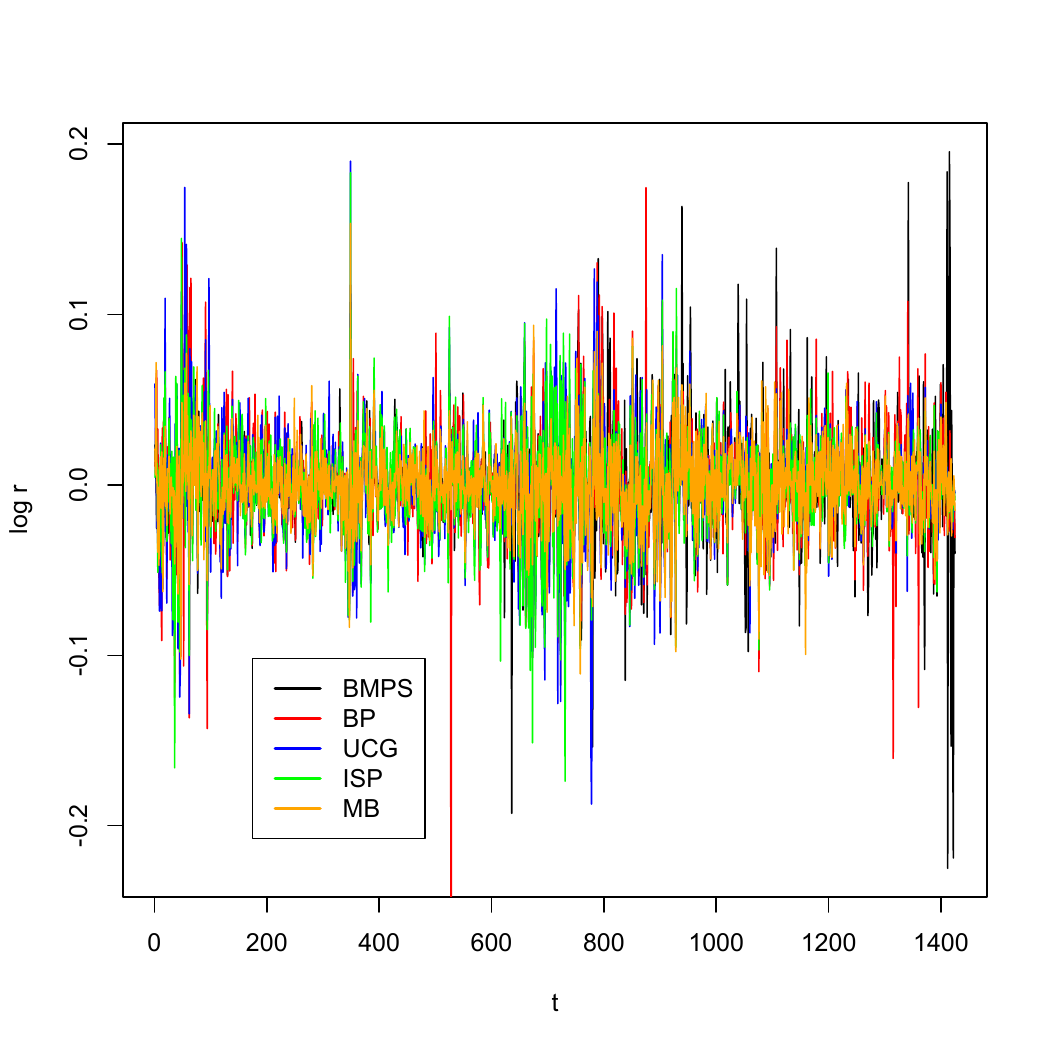}
}
\caption{Log-returns of Monte dei Paschi di Siena (BMPS), Banco Popolare (BP), Unicredit (UCG), Intesa-Sanpaolo (ISP), Mediobanca (MB) for weekdays from $01/07/2013$ to $30/06/2014$ .}
\label{returns}
\end{figure}

We now analyse a real dataset containing the log-returns FTSE-MIB of five Italian financial institutes (Monte dei Paschi di Siena,  Banco Popolare, Unicredit, Intesa-Sanpaolo and Mediobanca) by assuming that the log-returns for each bank may be modelled as a generalized autoregressive conditional heteroscedastic model with parameters $(1,1)$ and Student-\textit{t} innovations. Data refers to weekdays from $01/07/2013$ to $30/06/2014$; they are available on the web-page \textit{https://it.finance.yahoo.com} and are summarised in Figure \ref{returns}. 

This example shows the possibility, given by a copula representation of the multivariate distributions of the log-returns, to model the univariate distributions in a flexible and realistic way by considering fat tails, without concern on the existence or easiness of estimation of the multivariate version. GARCH models are known to suffer of the curse of dimensionality \citep{caporin:11}, however with the copula representation it is possible to separately model the marginal distributions in this way and then separately estimate the copula function. We will see in the following that the marginal modeling may have a limited impact on the estimation of the posterior distribution of the functional of interest of the dependence. 

The GARCH-\textit{t} model may be expressed via data augmentation \citep{geweke:93} as, for $t=1, \dots, T,$

\begin{align}
\label{eq:gracht}
y_t &= \varepsilon_t \sqrt{\frac{\nu-2}{\nu}\, \omega_t h_{t}}; \\
h_t &= \alpha_0+\alpha_1y_{t-1}^2+\beta h_{t-1}; \nonumber \\
\varepsilon_t &\sim \mathcal{N}(0,1); \nonumber \\
\omega_t &\sim IG\left(\frac{\nu}{2},\frac{\nu}{2}\right), \nonumber 
\end{align}

\noindent where $\alpha_0>0$, $\alpha_1,\beta>=0$, $\nu>2$ and $IG(a,b)$ denotes the inverse gamma distribution with shape parameter $a$ and scale parameter $b$. 
For each institute the posterior distribution of the model parameters ($\alpha_0,\alpha_1,\beta,\nu$) may be approximated by using the \erre package \texttt{bayesGARCH} \citep{ardia:10}.

There are several other models which may be used in this setting; for a comparison, we use the quantile distributions, which are defined in terms of their inverse cumulative distribution functions, which are functions of the quantiles of a standard normal distribution \citep{rayner2002numerical}. In particular, we use a \textit{g}-and-\textit{k}, given by

\begin{equation*}
Q(z; A,B,g,k) = A + B * \left( 1 + \frac{ 1-\exp{(-g z)}}{1+\exp{(-g z)}} \right) \left( 1 + z^2\right)^k z,
\end{equation*}

\noindent where $A$, $B$, $g$ and $k$ are the parameter of location, scale, skewness and kurtosis respectively and $z \sim \mathcal{N}(0,1)$. Here, we use the extension to time series data proposed by \cite{drovandi2011likelihood}, where $z_i$ follows a $MA(1)$ model 

\begin{equation}
z_i = \eta_i + \alpha \eta_{i-1}, \qquad i=1,\cdots,n
\end{equation}

\noindent where $n$ is the number of observations and $\eta_i \sim^{iid}N(0,1)$. Each $z_i$ is then diveded by $\sqrt{1+\alpha^2}$ to ensure it is marginally distribution as a standard normal. 
The most used approach to deal with this model, which is characterized by an intractable likelihood, is through approximate Bayesian computation \citep{allingham2009bayesian}, that is the approach we are using here. 
 
Once the marginal distributions are estimated, it is necessary to derive the pseudo-data to construct the copula. In this particular situation, the density function is analytically unavailable and so the distribution function. It is, therefore, possible to use a nonparametric approach to derive the pseudo-data, by using, for instance, a Pitman-Yor process prior as described in \cite{nieto2014bayesian}. This example shows the possibility to perform a separate analysis for the marginal distributions and the joint distribution. 

Once the pseudo-data have been derived, it is possible to apply Algorithm \ref{ABCOP}. Algorithm \ref{algo:ABCOP-logreturns} describes the several steps of implementation for the particular case of marginal GARCH-\textit{t} models, for the approximation of the posterior distribution of the first version of the multivariate Sperman's $\rho$, say $\rho_1$; for the tail dependence indices, it is possible to use a $\mathcal{U}nif(0,1)$ prior instead of the $\mathcal{U}nif(-1,1)$.  Without loss of generality, we have decided to describe the case where an equal number of simulations is chosen in the first step, in such a way that the posterior distributions of all the univariate marginals are approximated by samples of equal size.

\begin{algorithm}
\caption{ABSCop algorithm - GARCH \textit{t} model}
\label{algo:ABCOP-logreturns}

	\begin{enumerate}
		\vspace{0.1cm}
		\item STEP 1: Marginal estimation \\
		\vspace{0.1cm}
		Given a sample of log-returns $X=(X_1,X_2,\cdots,X_5)$
		 				where $X_j=(X_{1j},\cdots,X_{nj})$ for 
		 				$j=(1,\cdots,5)$ where each $X_j$ is assumed to follow a GARCH-\textit{t} model (as described in \eqref{eq:gracht} ) with parameters $(\alpha_{0j}, \alpha_{1j}, \beta_j, \nu_j)$
		 				
		\begin{algorithmic}[1]
			\For {$j=1,\cdots,5$ }
				\State Derive a posterior sample of $S$ values for $(\alpha_{0j}, \alpha_{1j}, \beta_j, \nu_j)$ approximating the marginal posterior $\pi(\alpha_{0j}, \alpha_{1j}, \beta_j, \nu_j|x_j)$; this may be achieved with any Monte Carlo approach, we have used the method proposed in \cite{ardia:10}.
				In this way, a $S \times 4$ matrix for any marginal distribution is obtained:
				$$
\lambda_j= \begin{pmatrix}
\alpha_{0j}^{(1)} & \alpha_{1j}^{(1)} & \beta_{j}^{(1)} & \nu_{j}^{(1)} \\
\alpha_{0j}^{(2)} & \alpha_{1j}^{(2)} & \beta_{j}^{(2)} & \nu_{j}^{(2)} \\
\dots  & \dots  & \dots & \dots  \\
\alpha_{0j}^{(S)} & \alpha_{1j}^{(S)} & \beta_{j}^{(S)} & \nu_{j}^{(S)} \\
\end{pmatrix} ,
$$	
\noindent for $j=1,\cdots,5.$			
			\EndFor
		\end{algorithmic}
		
		\item STEP 2: Joint estimation

		\begin{algorithmic}[1]
			\For {$b=1,\cdots,B$}
				\State Draw $\rho^{(b)} \sim \mathcal{U}nif(-1,1)$
				\For {$s=1,\cdots,S$}
					\State Pick the $s$-th row of each matrix $\lambda_j$, i.e. $\left(\alpha_{0j}^{(s)},\alpha_{1j}^{(s)}, \beta_{j}^{(s)}, \nu_{j}^{(s)}\right)$ for $j=1,\cdots,5$.		
					\State Derive a matrix of uniformly distributed pseudo-data $u_{ij}=F_j(x_{ij};\alpha_{0j}^{(s)},\alpha_{1j}^{(s)}, \beta_{j}^{(s)}, \nu_{j}^{(s)})$, where $F_j(\cdot;\alpha_{0j}^{(s)},\alpha_{1j}^{(s)}, \beta_{j}^{(s)}, \nu_{j}^{(s)})$ is the distribution function of a GARCH-\textit{t} model with parameters $\left(\alpha_{0j}^{(s)},\alpha_{1j}^{(s)}, \beta_{j}^{(s)}, \nu_{j}^{(s)}\right)$:
				$$
u^{(s)}= \begin{pmatrix}
u_{11}^{(s)} & u_{12}^{(s)} & \dots & u_{15}^{(s)} \\
u_{21}^{(s)} & u_{22}^{(s)} & \dots & u_{25}^{(s)} \\
\dots  & \dots  & u_{ij}^{(s_j)} & \dots  \\
u_{n1}^{(s)} & u_{n2}^{(s)} & \dots & u_{n5}^{(s)} \\
\end{pmatrix} .
$$		
	
				\State Compute a nonparametric estimate of the Spearman's $\rho_1$: 
				$$ \hat{\rho}_{1n}^{(s)}=h(5)\left\{\frac{2^5}{n}\sum_{i=1}^n \prod_{j=1}^5 (1-u^{(s)}_{ij}) -1 \right\}$$
				where $h(5) = (5+1)/ \{2^5-(5+1)\}$
				
				\State Compute $L_{BEL}(\rho_1^{(b)};u^{(s)})=\omega_{bs}$
				
				\EndFor
		
				\State \Return A sample of size $B$ of values of $\rho_1$ from the prior distribution and a $B \times S$ matrix of weights. 
				
				\State Compute the row average weight $\bar{\omega}_b=\frac{1}{S} \sum_{i=1}^S \omega_{bs}$
				
			\EndFor
			\State Sample with replacement $(\rho_1^{(b)}\bar{\omega}_b)$, $b=1,\cdots,B$.
		\end{algorithmic}

	\end{enumerate}

	\textbf{Output:} a sample of size $B$ of values approximately from the posterior distribution of $\rho_1$.
\end{algorithm}

\begin{figure}
\centerline{
\includegraphics[width=28pc,height=15pc]{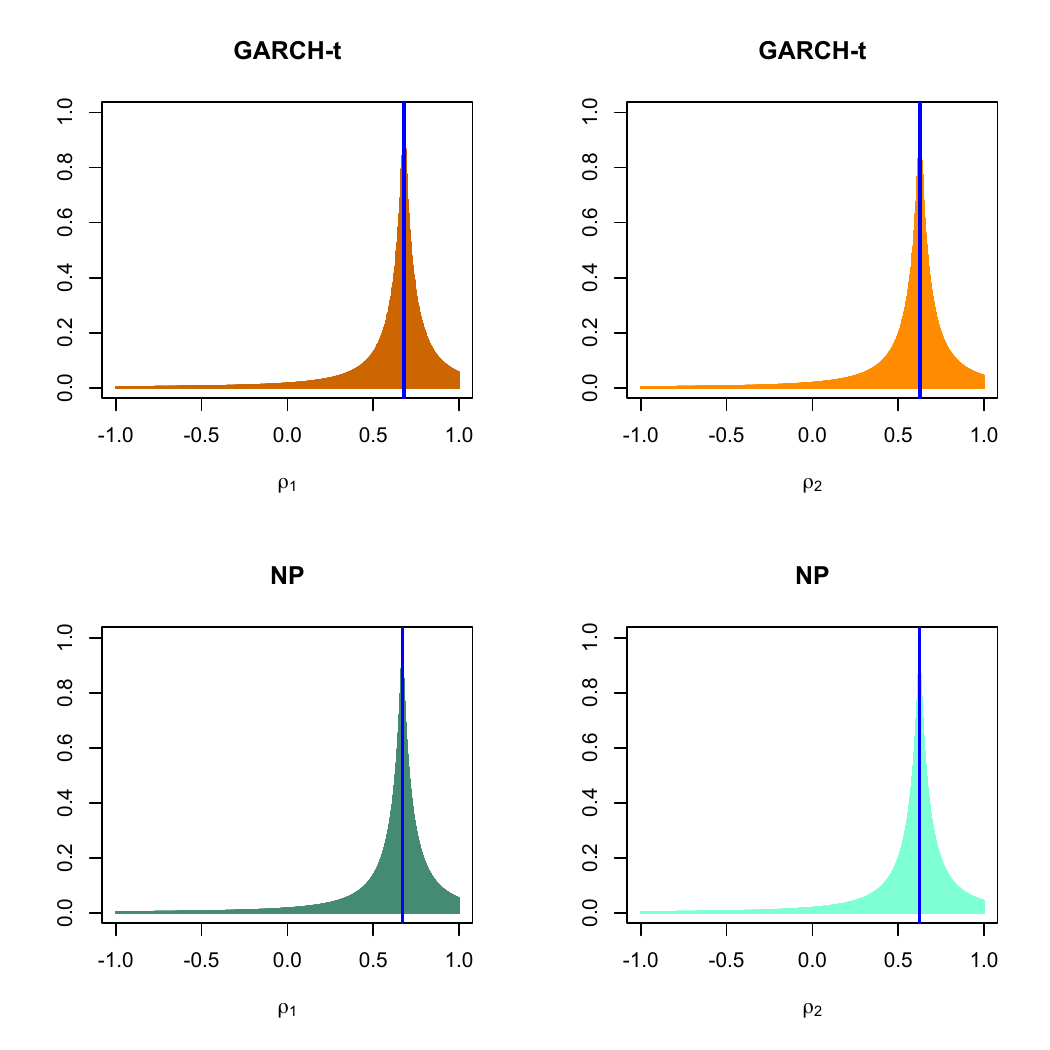}
}
\caption{Approximation of the posterior distribution of the multivariate Spearman's $\rho_1$ (left) and $\rho_2$ (right) for the log-returns of the investments of five Italian institutes based on $10,000$ simulations. On the top, there are the approximation obtained by assuming a GARCH-\textit{t}(1,1) model, while at the bottom there are the approximation obtained by using a nonparametric procedure in the marginal estimation. The blue vertical lines represent the frequentist estimates.}
\label{garcht_multi}
\end{figure}

Figure \ref{garcht_multi} shows the results of the Bayesian procedure based on Algorithm \ref{ABCOP} for $\rho_1$ and $\rho_2$ as defined in Section \ref{sec:multianalysis}. The approximated Bayesian posterior means are $0.604$ and $0.600$ for $\rho_1$ with the parametric and the nonparametric procedure respectively and $0.559$ and $0.561$ for $\rho_2$; the posterior distribution are centred around the frequentist estimates, however they provide a quantification of the uncertainty that is not available with the frequentist procedure (as already shown in Section \ref{sec:multianalysis}). The impact of the choice of the estimation procedure at the marginal step does not seem to have a great impact on the results.  

\begin{figure}
\centerline{
\includegraphics[width=28pc,height=15pc]{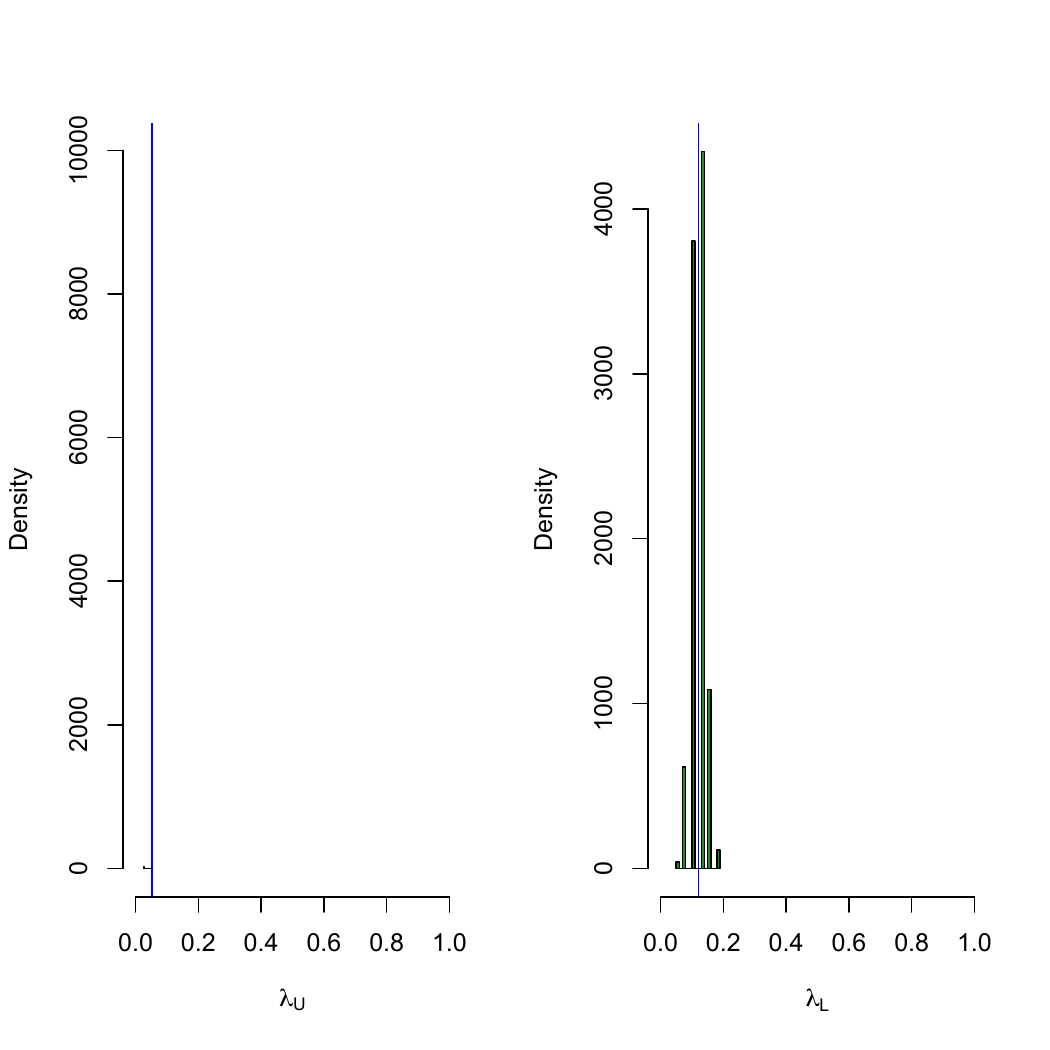}
}
\caption{Approximation of the posterior distribution of the multivariate tail dependence indices, $\lambda_U$ (left) and $\lambda_L$ (right) for the log-returns of the investments of five Italian institutes based on $10000$ simulations. The blue lines are the frequentist estimates.}
\label{garcht_multi_taildep}
\end{figure}

Figure \ref{garcht_multi_taildep} shows the corresponding results for the multivariate tail dependence indices as defined in Section \ref{sec:multianalysis}. In this case, the posterior distributions are strongly concentrated around small values, the posterior for $\lambda_U$ is strongly concentrated around $0$ and the posterior for $\lambda_L$ is concentrated around $0.12$. 

The gist of this example is to emphasize the role of the Bayesian approach and the copula representation in the quantification process of tail codependence among different series which would be very hard by simply looking at Figure \ref{returns}.

\section{Discussion}

This paper describes a novel method for obtaining the posterior distribution of a quantity of interest in cases where either the model is only partially specified or the computation or the evaluation of the complete likelihood is too costly. 

In particular, we have considered the case of copula models, although extensions to the general semiparametric approach in a Bayesian framework are easy to consider. 

The proposed method is deliberately approximated, since it avoids a complete specification of the statistical model. This may be extremely useful in applications where the user is only interested in a particular aspect of the data, for example in particular summaries of the dependence structure. In these situations, the introduction of any further parametric assumption would generally be difficult to verify and defend and it would presumably introduce a lack of robustness. 

Finally, our approach provides a natural quantification of the uncertainty of the estimates of common measures of dependence in copula theory, in contrast with standard available methods.

\section*{Supplementary Material}

\subsection*{Appendix A: Tail Dependence}
\label{sec:appA}

This appendix show more simulated examples for the study of tail dependence, in particular simulations from a Frank copula (with $\theta=3.45$) and from a Gumbel copula (with $\theta=2$).

\begin{figure}
\centerline{
\includegraphics[width=28pc,height=15pc]{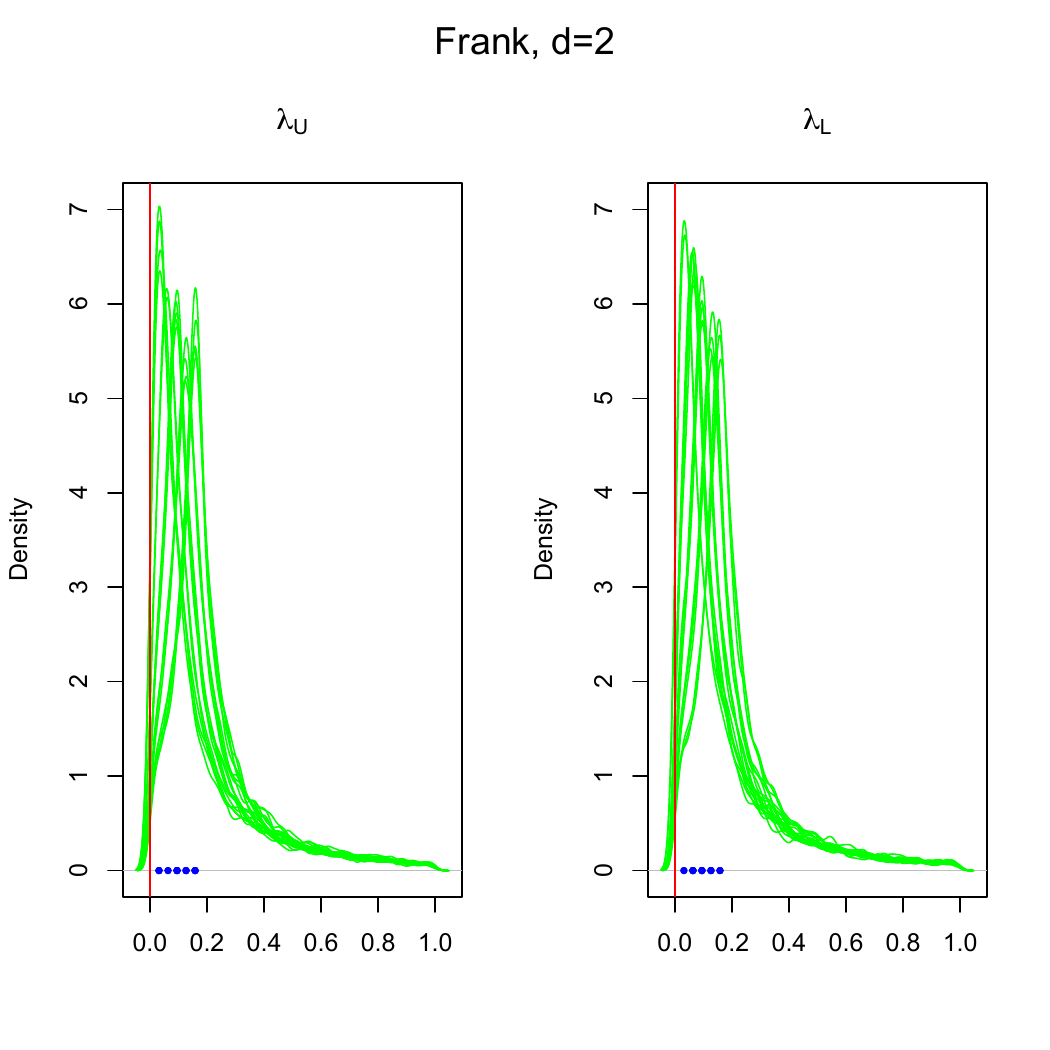}
}
\caption{Comparison between frequentist (blue) and Bayesian (green) estimates for $\lambda_U$ (above) and $\lambda_L$ (below). $20$ out of $500$ simulations from a Frank copula with $\theta=3.45$:($n=1000$); the circles represent the frequentist point estimates, the lines represent the posterior distributions. The true values are $\lambda_U^{true}=0$ and $\lambda_L^{true}=0$ (red lines).}
\label{fig:lambda_frank}
\end{figure}

\begin{figure}
\centerline{
\includegraphics[width=28pc,height=15pc]{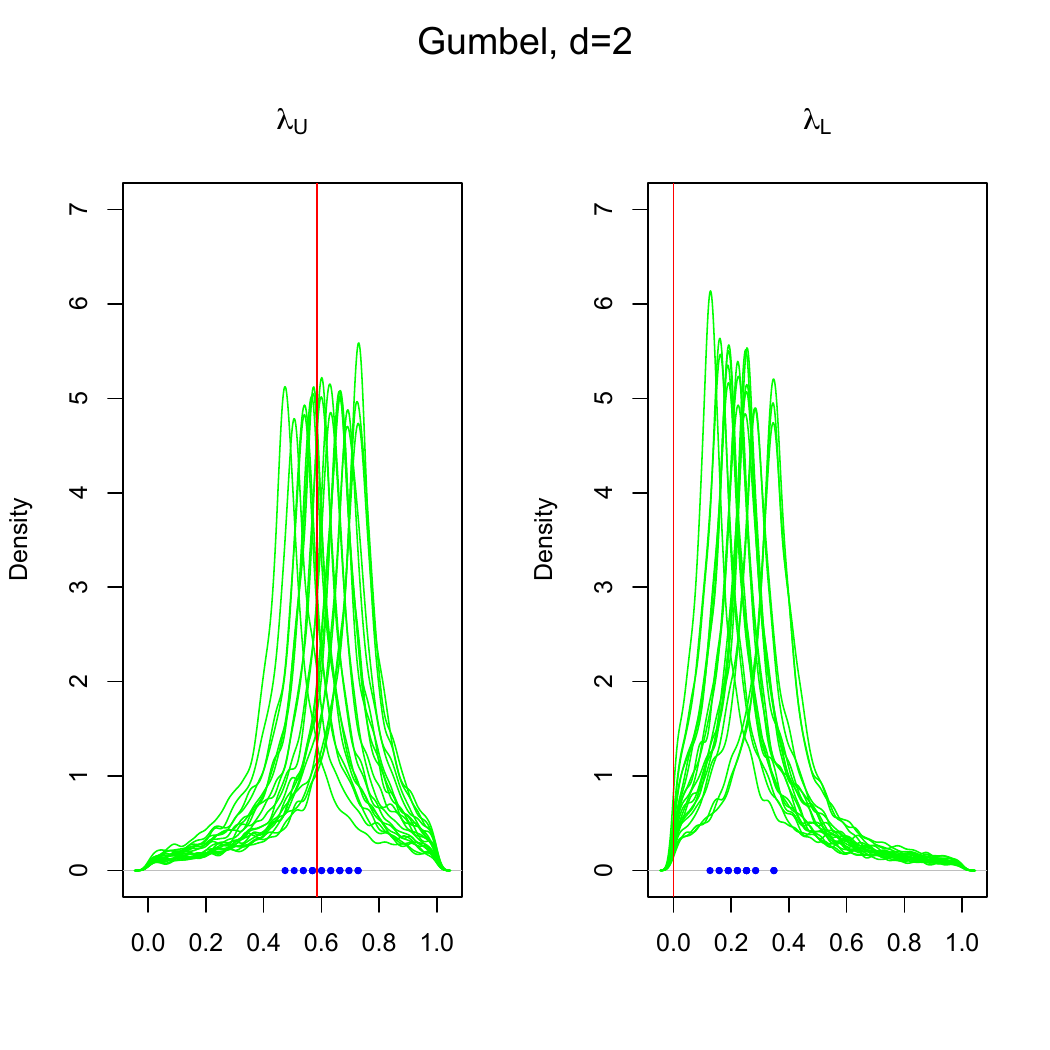}
}
\caption{As in Figure \ref{fig:lambda_frank}, for a Gumbel copula with $\theta=2$; the true values, $\lambda_U^{true}=2-2^{\frac{1}{\theta}}$ and $\lambda_L^{true}=0$ are the red horizontal lines.}
\label{fig:lambda_gumbel}
\end{figure}

\subsection*{Appendix B: Multivariate Analysis}
\label{sec:appB}

\subsubsection{Multivariate Spearman's $\rho$}

\begin{figure}
\centerline{
\includegraphics[width=28pc,height=15pc]{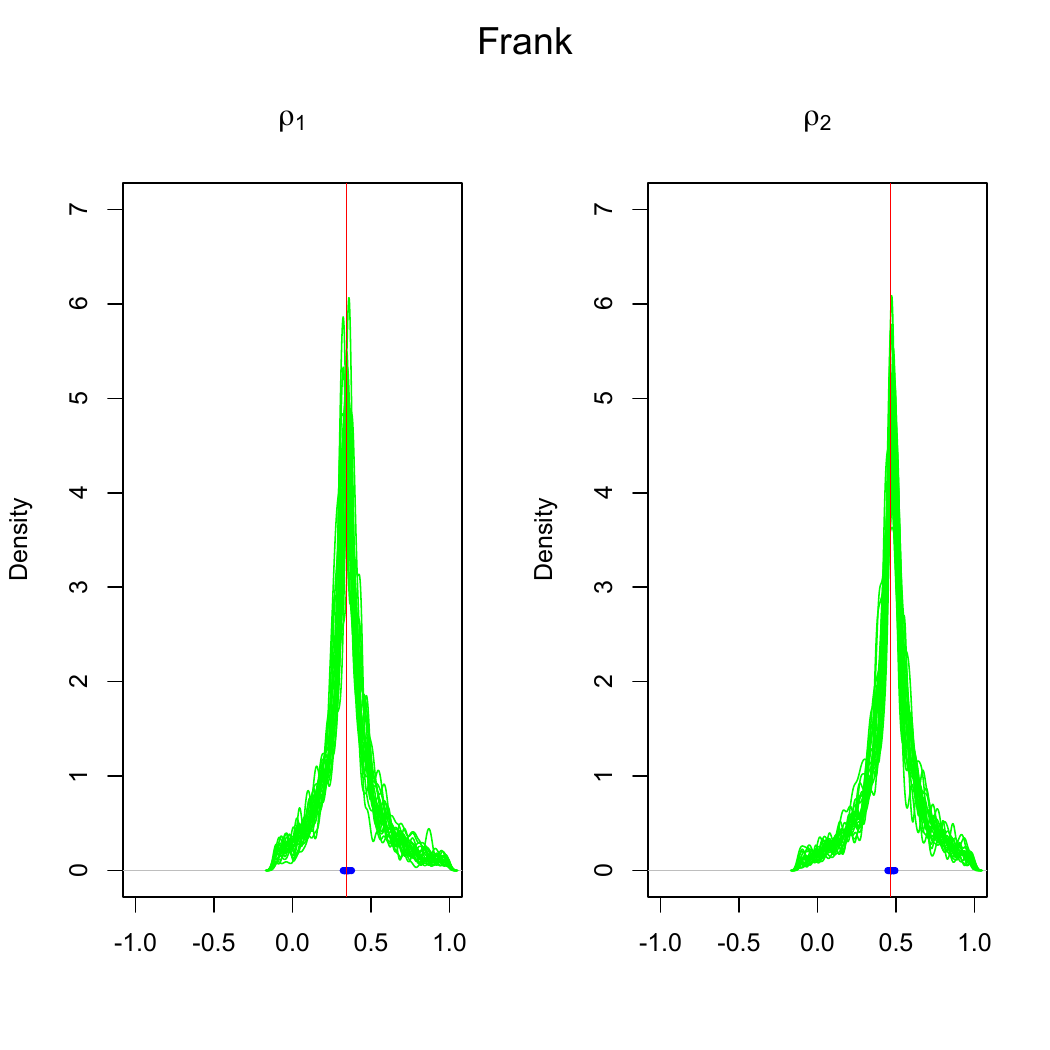}
}
\caption{Comparison between frequentist (blue) and Bayesian (green) estimates of $\rho_1$ (above) and $\rho_2$ (below). $20$ out of $500$ experiments with simulation from a Frank copula with $\theta=3.45$ ($n=1000$); the true values are the red horizontal lines, the circles represent the point estimates and the lines represent the interval estimates. The frequentist confidence intervals are not visible because of a too small estimate of the variance.}
\label{fig:rho6_frank}
\end{figure}

\begin{figure}
\centerline{
\includegraphics[width=28pc,height=15pc]{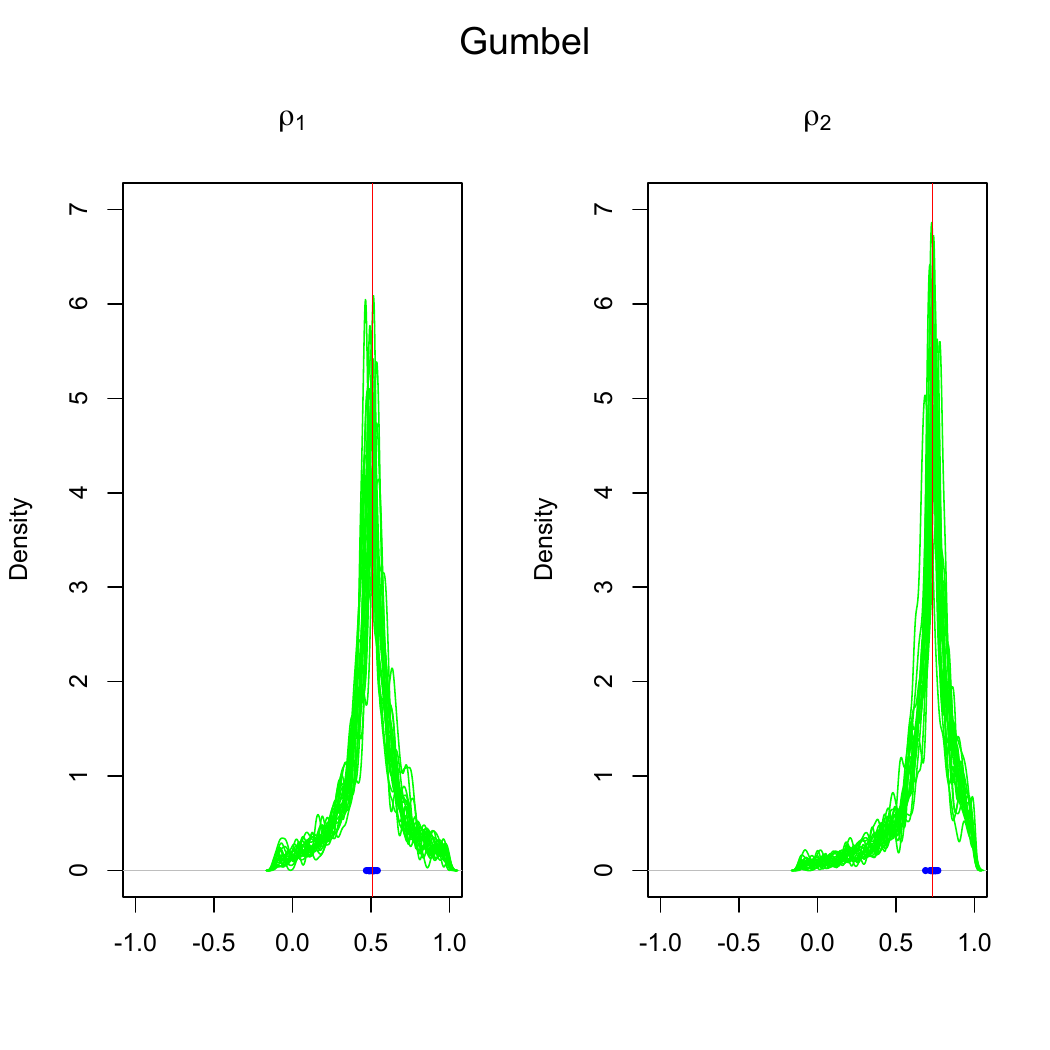}
}
\caption{As in Figure \ref{fig:rho6_frank}, simulations from a Gumbel copula with $\theta=2$.
}
\label{fig:rho6_gumbel}
\end{figure}

\begin{figure}
\centerline{
\includegraphics[width=28pc,height=15pc]{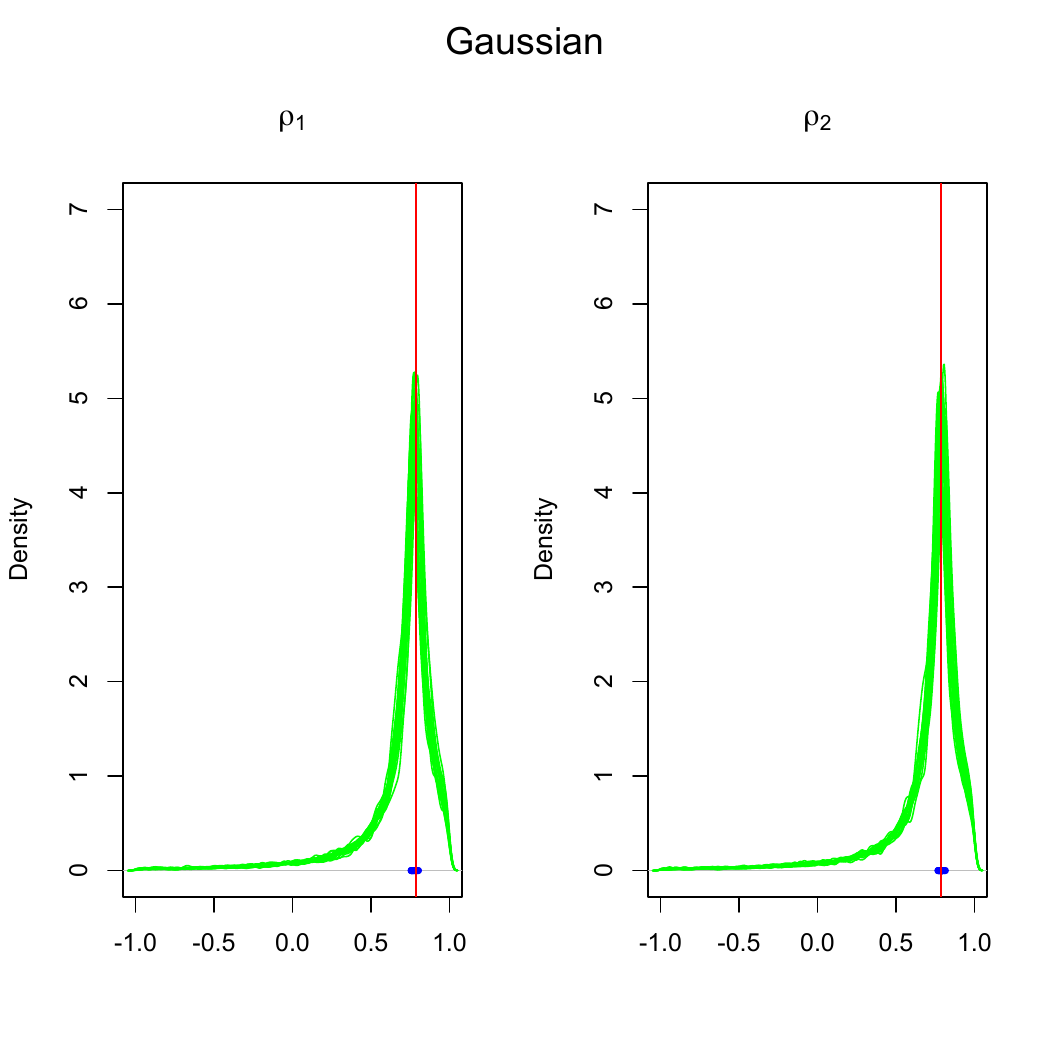}
}
\caption{As in Figure \ref{fig:rho6_frank}, simulations from a Gaussian copula with $\rho_1=\rho_2=0.8$.
}
\label{fig:rho6_gaussian}
\end{figure}

Figures \ref{fig:rho6_frank}, \ref{fig:rho6_gumbel} and \ref{fig:rho6_gaussian} show some additional examples for the frequentist confidence intervals and the corresponding Bayesian equal tails intervals (of level $0.95$) for simulations from 

\begin{itemize}
\item a Frank copula with $\rho_1=0.343$ and $\rho_2=0.465$,
\item a Gumbel copula with $\rho_1=0.511$ and $\rho_2=0.734$,
\item a Gaussian copula with $\rho_1=\rho_2=0.800$.
\end{itemize}

\subsubsection{Multivariate tail dependence}

\begin{figure}
\centerline{
\includegraphics[width=28pc,height=15pc]{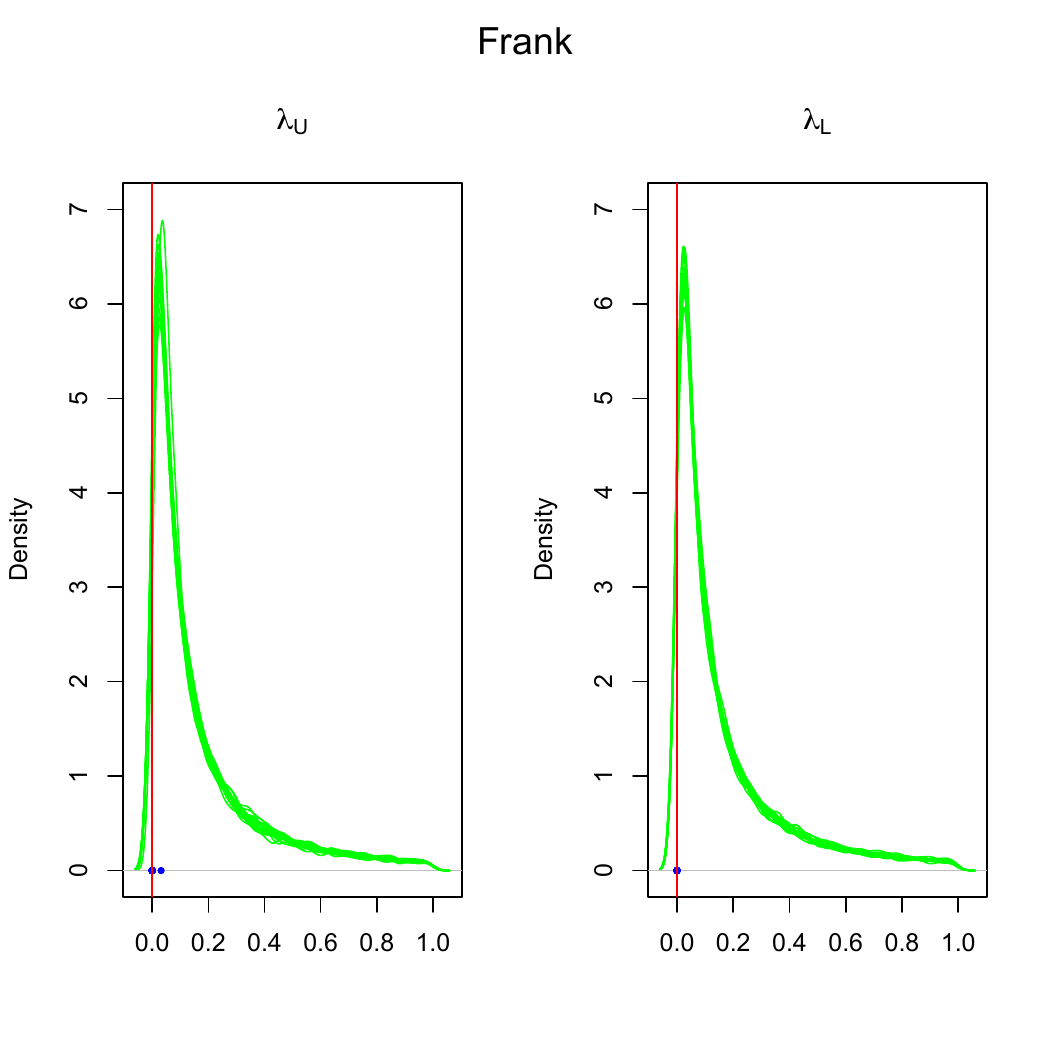}
}
\caption{Approximated posterior distributions of $\lambda_U$ (left) and $\lambda_L$ (right) for simulations from a six-dimension Frank copula with $\theta=3.45$ in $20$ out $500$ experiments. The blue points are the frequentist point estimates, the red lines are the true values, $\lambda_U^{true}=0$ and $\lambda_L^{true}=0$.}
\label{fig:multiTD_frank}
\end{figure}

\begin{figure}
\centerline{
\includegraphics[width=28pc,height=15pc]{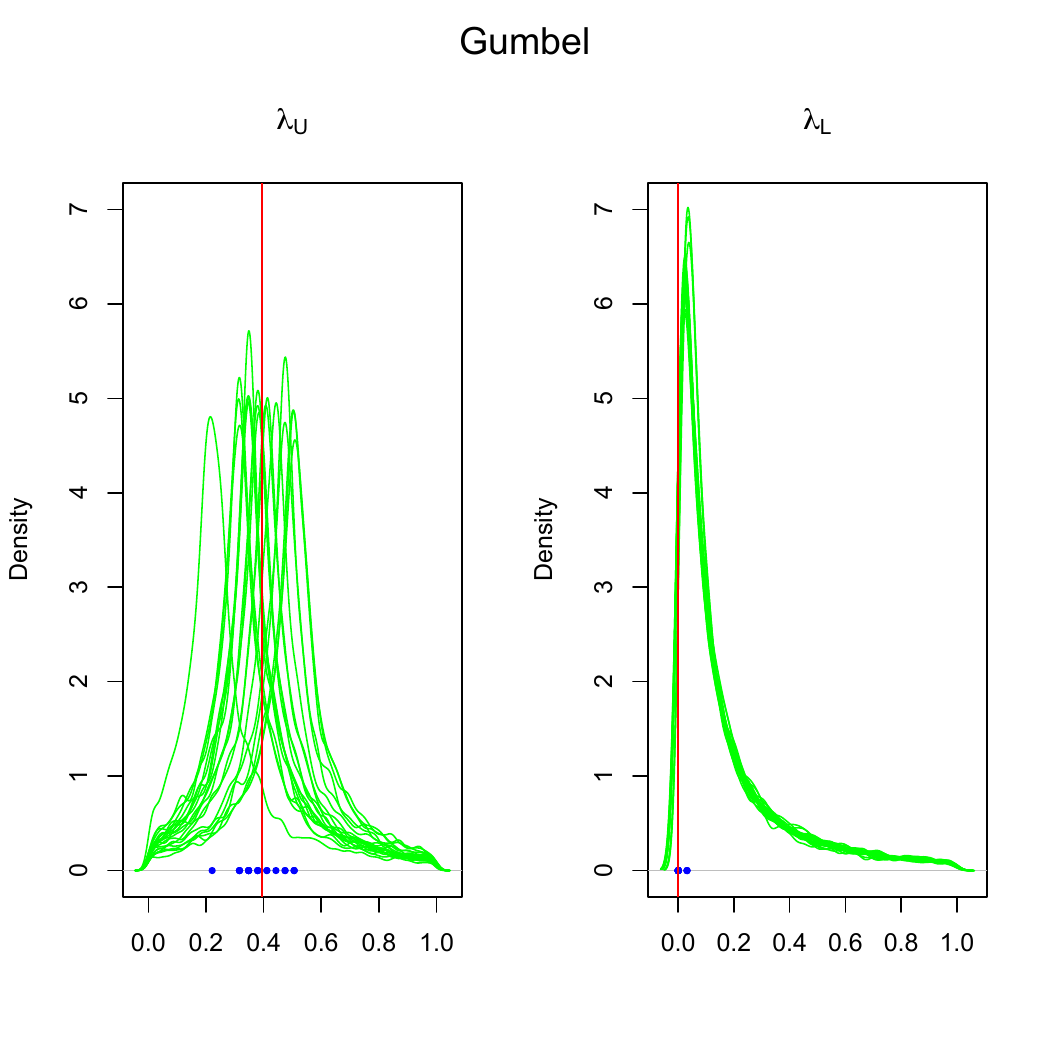}
}
\caption{Approximated posterior distributions of $\lambda_U$ (left) and $\lambda_L$ (right) for simulations from a six-dimension Gumbel copula with $\theta=2$ in $20$ out $500$ experiments. The blue points are the frequentist point estimates, the red lines are the true values, $\lambda_U^{true}=0.395$ and $\lambda_L^{true}=0$.}
\label{fig:multiTD_gumbel}
\end{figure}

Figure \ref{fig:multiTD_frank} and \ref{fig:multiTD_gumbel} show some additional examples for the frequentist points estimates and the corresponding Bayesian posterior distribution of $\lambda_U$ and $\lambda_L$ for simulations from 

\begin{itemize}
\item a Frank copula with $\theta=3.45$, $\lambda_U=0$ and $\lambda_L=0$,
\item a Gumbel copula with $\theta=2$, $\lambda_U=0.395$ and $\lambda_L=0$.
\end{itemize}

\end{document}